\documentclass[10pt,reqno]{article}
\usepackage{amssymb, amsmath, amsfonts}
\usepackage[english]{babel}
\usepackage{bm}
\usepackage{graphicx, epsfig}
\usepackage{epstopdf}
\newtheorem{theorem}{Theorem}[section]
\newtheorem{cor}[theorem]{Corollary}
\newtheorem{lemma}[theorem]{Lemma}
\newtheorem{prop}[theorem]{Proposition}
\newtheorem{remark}[theorem]{Remark}

\newcommand{\nm}{\noalign{\smallskip}}

\def\ep{\epsilon}

\newcommand{\Bp}{\mathbf{p}}

\newcommand{\Bc}{\mathbf{c}}

\newcommand{\Om}{\Omega}
\newcommand{\Bx}{\mathbf{x}}
\newcommand{\By}{\mathbf{y}}

\newcommand{\RR}{\mathbb{R}}
\newcommand{\CC}{\mathbb{C}}

\newcommand{\Scal}{\mathcal{S}}
\newcommand{\Dcal}{\mathcal{D}}

\newcommand{\p}{\partial}

\newcommand{\pd}[2]{\frac {\p #1}{\p #2}}
\newcommand{\ds}{\displaystyle}
\newcommand{\eqnref}[1]{(\ref {#1})}
\newcommand{\pf}{\medskip \noindent {\sl Proof}. \ }
\newcommand{\qed}{\hfill $\Box$ \medskip}
\newcommand{\beq}{\begin{equation}}
\newcommand{\eeq}{\end{equation}}

\newcommand{\la}{\langle}

\numberwithin{equation}{section}
\numberwithin{figure}{section}

\begin{document}

\title{Characterization of the gradient blow-up of the solution to the conductivity equation in the presence of adjacent circular inclusions\thanks{\footnotesize
This research was supported by the Korean Ministry of Science, ICT and Future Planning through NRF grants No. 2013012931 and No. 2013003192.}}

\author{Mikyoung
Lim\thanks{\footnotesize Department of Mathematical Sciences,
Korea Advanced Institute of Science and Technology, Daejeon
305-701, Korea (mklim@kaist.ac.kr, shyu@kaist.ac.kr).} \and
Sanghyeon Yu\footnotemark[2]}

\maketitle

\begin{abstract}
We consider the conductivity problem in the presence of adjacent circular inclusions having arbitrary constant conductivity.
 When two inclusions get closer and their conductivities degenerate to zero or infinity, the gradient of the solution can be arbitrary large.
We characterize the gradient blow-up by deriving an explicit formula for the singular term of the solution in terms of the Lerch transcendent function. This derivation is valid for inclusions having arbitrary constant conductivity. We illustrate our results with numerical calculations.
\end{abstract}



\section{Introduction}
We consider the blow-up phenomena of the gradient of the solution to the conductivity problem when two conductors are closely located to each other in $\RR^2$. Let $B_1$ and $B_2$ be two disks with conductivity $k_1$ and $k_2$, respectively, embedded in the background with conductivity 1. The conductivities $k_1$ and $k_2$ are assumed to be $0<k_1, k_2\neq1<\infty$. Let $\sigma$ denote the conductivity distribution, \textit{i.e.},
\beq\label{sigma:def}
\sigma=k_1\chi(B_1)+k_2\chi(B_2)+\chi(\RR^2\setminus (B_1\cup B_2),
\eeq
where $\chi$ is the characteristic function.
Consider the following conductivity equation:
\beq\label{cond_eqn}
\begin{cases}
\ds\nabla\cdot\sigma\nabla u=0\quad&\mbox{in }\RR^2, \\
\ds u(\Bx) - H(\Bx)  =O(|\Bx|^{-1}) \quad&\mbox{as } |\Bx| \to \infty,
\end{cases}
\end{equation}
where $H$ is a given entire harmonic function in $\RR^2$. On the boundary of inclusions, the solution $u$ to \eqnref{cond_eqn} satisfies
\beq\label{u:trans}\ds u\bigr|^-_{\p B_j}=u\bigr|^+_{\p B_j} \mbox{ and } k_j\pd{u}{\nu}\Bigr|^-_{\p B_j}=\pd{u}{\nu}\Bigr|^+_{\p B_j},\ j=1,2.\eeq
The superscripts $+$ and $-$ denote the limit from inside and outside $\p B_j$, respectively.
Letting
$$\ep:=\mbox{dist}(B_1,B_2),$$
we estimate $\nabla u$ in terms of $\ep$ as $\ep$ tends to 0,
where the shape of $B_1$ and $B_2$ are fixed.
This problem arises in relation with the computation of electromagnetic fields in the presence of fibers and the stress in composite materials.

For the conductivities finite and strictly positive, it was proved by Li and Vogelius \cite{LV} that $|\nabla u|$ is bounded independently of $\ep$, and Li and Nirenberg \cite{LN} have extended this result to elliptic systems.  Bonnetier and Vogelius showed in \cite{BV} that $|\nabla u|$ is bounded for circular touching inclusions.
However, if $k_1,k_2$ degenerate to $\infty$, then the gradient may blow-up as $\ep$ tends to 0. The generic rate of gradient blow-up is $\ep^{-1/2}$ in two dimensions \cite{AKLLL, AKLLZ, AKL, bab, BLY, BT, BC,keller, Y, Y2}, while it is $|\ep\log\ep|^{-1}$ in three dimensions \cite{BLY, BLY2,KLY2, LY}.
Ammari, Kang and Lim \cite{AKL} and Ammari, Kang, Lee, Lee and Lim \cite{AKLLL} obtained an optimal bound for $|\nabla u|$ when $B_1$ and $B_2$ are disks in $\RR^2$ for $0<k_1, k_2\neq1<\infty$. Yun \cite{Y, Y2} obtained the blow-up rate for the perfectly conducting general shaped inclusions.
In three dimensions, the blow-up rate was obtained by Bao, Li and Yin \cite{BLY, BLY2} for perfect conductors of general shape. For perfect conductors of spherical shape, Lim and Yun \cite{LY} obtained an optimal bound for $|\nabla u|$, and Kang, Lim and Yun \cite{KLY2} derived an asymptotic of $\nabla u$. The insulating case has the same blow-up rate as the perfectly conducting case in two dimensions. However, it is still open problem to clarify whether the gradient of $u$ may blow-up or not for the insulating cases in three dimensions.

In this paper, we are interested in the dependence of the singular part of $u$ on the conductivity $k_1$ and $k_2$ as well as $\ep$. To state the related results in detail, let us introduce notations.  For $j=1,2$, let $B_j$ be the disk with radius $r_j$ centered at $\Bc_j$ and $R_j$ be the reflection with respect to $B_j$, \textit{i.e.},
\beq\notag R_j(\mathbf{x})=\frac{r_j^2(\Bx-{\Bc}_j)}{|\Bx-{\Bc}_j|^2}+{\Bc}_j,\quad j=1,2.\eeq
Then the combined reflections $R_1\circ R_2$ and $R_2\circ R_1$ have unique fixed points, say $\Bp_1$ and $\Bp_2$, respectively. It is easy to show that, for $j=1,2$,
\beq\notag\mathbf{p}_j=\Bigr((-1)^jr_*\sqrt{\ep}+O(\ep)\Bigr)\mathbf{n}+\mathbf{p},\ \mathbf{n}=\frac{\Bp_2-\Bp_1}{|\Bp_2-\Bp_1|},\eeq
where $\Bp$ is the middle point of the shortest line segment connecting $\p B_1$ and $\p B_2$.
We also set \beq r_*=\sqrt{{(2r_1r_2)}/{(r_1+r_2)}},\
\tau=\tau_1\tau_2,\mbox{ and }
\label{tau:def}\tau_j=(k_j-1)/(k_j+1),\ j=1,2.\eeq

In \cite{AKLLZ}, it was shown that the gradient of $u$ may blow-up only when the linear term of $H$ is nonzero. Hence $u$ can be decomposed as
\beq\label{usur}u= u_s+u_r,\eeq
where $u_s$ is the solution to \eqnref{cond_eqn} with $H$ replaced by $\tilde{H}(\Bx)=\nabla H(\Bp)\cdot(\Bx-\Bp)$ and the gradient of $u_r$ stays bounded regardless of $\epsilon$ in a bounded domain. Using the representation of $u$ in terms of the single-layer potential, one can actually show that the gradient of $(u_r-H+\tilde{H})$ is bounded in $\RR^2$.
The following estimates of $|\nabla u_s|$ were derived in \cite{AKLLL, AKLLZ, AKL}: if $k_1,k_2>1$, then for any bounded set containing the inclusions, there are positive constants $C_1$ and $C_2$ independent of $k_1,k_2,r_1,r_2,\ep$ such that
\begin{align}\label{gblowup}
\ds\bigr|\nabla u_s\bigr|^+(\Bx_j)&\geq\frac{C_1|\nabla H(\Bp)\cdot \mathbf{n}|}{1-\tau+\frac{r_*}{\min(r_1,r_2)}\sqrt\ep},\ j=1,2,\\[2mm]\ds
\label{gblowup2}\bigr\|\nabla u_s\bigr\|_{L^\infty(\Om)}&\leq\frac{C_2|\nabla H(\Bp)\cdot \mathbf{n}|}{1-\tau+\frac{r_*}{\max(r_1,r_2)}\sqrt\ep},\end{align}
where $\Bx_j$, $j=1,2$, is the point on $\p B_j$ closest to the other disk.
For $0<k_1,k_2<1$, we have \eqnref{gblowup} and \eqnref{gblowup2} with $\mathbf{n}$ replaced by $\mathbf{t}$ and $k_j$ by $1/k_j$, $j=1,2$,
where $\mathbf{t}$ is the rotation of $\mathbf{n}$ by $\pi/2$-radians.

For the case of perfectly conducting, $k_1=k_2=\infty$, the singular term $u_s$ can be explicitly characterized using the singular function $h$ defined as the solution to
\beq\label{h:eqn}
\quad \left\{
\begin{array}{ll}
\ds\Delta h=0 \quad& \mbox{in } \mathbb{R}^2 \setminus \overline{B_1 \cup B_2}, \\
\ds h= \mbox{constant}\quad& \mbox{on }\p B_j, j=1,2,\\
\nm
\ds \int_{\p B_j} \pd{h}{\nu^{(j)}} ~ds=(-1)^{j+1},\ \quad& j=1,2,\\
\ds h(\Bx)=O(|\Bx|^{-1}) \quad&\mbox{as } |\Bx| \to \infty,
\end{array}
\right.
\end{equation}
where $\nu^{(j)}$ is outward unit normal to $\p B_j$. Remind the circle of Apollonius:
for a disk $B_r(\Bc)$ with radius $r$ centered at $\Bc$, the circle $\p B_r(\Bc)$ is the locus of $\Bx$ satisfying the ratio condition
\beq\label{apollo:1202}
\frac{|\Bx-\Bp|}{|\Bx-R(\Bp)|}=\frac{|\Bp-\Bc|}{r},\quad\mbox{for a fixed }\Bp\notin\overline{B_r(\Bc)},\eeq
where $R$ is the reflection with respect to $B_r(\Bc)$. Since $R_1(\Bp_2)=\Bp_1$ and $R_2(\Bp_1)=\Bp_2$, we have
\beq\label{h:apo}\ds 2\pi h(\Bx)= \ln|\Bx-\Bp_1|-\ln|\Bx-\Bp_2|.\eeq In \cite{KLY}, it was shown that  \beq \label{charbup}
u(\Bx)= 2\pi c_\mathbf{n} h(\Bx)+u_r(\Bx)\quad\mbox{with }c_\mathbf{n}= r_*^2\bigr(\nabla H(\Bp)\cdot \mathbf{n}\bigr),\eeq where the gradient of $u_r$ is bounded regardless of $\epsilon$ in any bounded set.
The solution $h$ to \eqnref{h:eqn} plays an essential role in the characterization of the blow-up feature not only for the circular inclusions. By investigating $h$, it was derived the optimal bounds of the gradient of $u$ for convex domains in two dimensions \cite{Y,Y2} and for balls in three dimensions \cite{LY, LY2}. In \cite{ACKLY}, for convex inclusions in two dimensions, the blow-up of $\nabla u$ was characterized using $h$ corresponding to the disk osculating to inclusions. It is worth to mention that, as shown in \cite{ACKLY}, $h$ is an eigenfunction corresponding to eigenvalue 1/2 of a Neumann-Poincar\'e operator.

The purpose of this paper is to generalize \eqnref{charbup} to the case of circular inclusions with arbitrary constant conductivity. More precisely speaking, we characterize the singular term of $u$ when two circular inclusions with the conductivity $0<k_1, k_2\neq1<\infty$ gets closer. To do this, we expand the solution $u$ in bipolar coordinate system $(\xi,\theta)$ with poles located at $\Bp_1$ and $\Bp_2$. In other words, for $\Bx=(x,y)$, the coordinates are defined as
\beq\label{xitheta1}
e^{\xi+i\theta}=\frac{z-\tilde{\Bp}_1}{\tilde{\Bp}_2-z},\quad z=x+iy,
\eeq
where $(\Re\{\tilde{\Bp}_j\},\Im\{\tilde{\Bp}_j\})=\Bp_j$, $j=1.2$.
It is worth to note that
\begin{equation}\notag
\xi(\Bx)=2\pi h(\Bx).
\eeq
Applying \eqnref{apollo:1202}, the first coordinate
$\xi$ takes the constant value $(-1)^j\xi_j$  on $\p B_j$ with
\beq\label{def:xi12}
\xi_1:=\ln|\Bp_2-\Bc_1|-\ln( r_1)\mbox{ and }\xi_2:=\ln|\Bp_1-\Bc_2|-\ln( r_2).\eeq

To state the main theorems we also define a function
\begin{align}
&q(\Bx;\beta,\tau_1,\tau_2)\nonumber\\\label{a_ep}
&: =  \frac{1}{2}
\begin{cases}
\ds (\tau_1+\tau) L\bigr(e^{-(\xi+i\theta)-2\xi_1};\beta\bigr)-(\tau_2+\tau)L\bigr(e^{(\xi+i\theta)-2\xi_2};\beta\bigr) \  &\mbox{in }\mathbb{R}^2 \setminus \overline{(B_1 \cup B_2)},\\[1mm]
 \ds   (\tau_1+\tau) L\big(e^{\xi-i\theta};\beta\big) -(\tau_2+\tau)L\bigr(e^{(\xi+i\theta)-2\xi_2};\beta\bigr)\quad &\mbox{in }B_1,\\[1mm]
 \ds (\tau_1+\tau)  L\bigr(e^{-(\xi+i\theta)-2\xi_1};\beta\bigr) -(\tau_2+\tau)L\big(e^{-\xi+i\theta};\beta\big)\quad &\mbox{in }B_2,
  \end{cases}
\end{align}
where \beq \label{parameter_beta}
\beta =\frac{r_*(-\ln\tau)}{4\sqrt{\epsilon}},\eeq
$$
L(z;\beta):=-\int_0^1 \frac{zt^\beta}{1+zt}\ dt,\quad z\in\CC,\ |z|<1.$$

The followings are the main results in this paper. Theorem \ref{main_thm_u_y} can be proved similarly as Theorem \ref{main_thm_u_x}.
\begin{theorem}\label{main_thm_u_x}
For $k_1,k_2>1$, the solution $u$ to \eqnref{cond_eqn} can be expressed as
\beq\label{eqn:main_thm}
u(\Bx) = c_{\mathbf{n}} \Re\{q(\Bx;\beta,\tau_1,\tau_2)\}+ H(\Bx)+ u_b(\Bx),
\eeq
where $c_\mathbf{n}$ is in \eqnref{charbup} and there is a constant $C$ independent of $\ep$, $k_1$ and $k_2$
such that \beq\notag \|\nabla u_b\|_{L^\infty(\RR^2)}\leq C.\eeq
Here $|\nabla u_b|$ on $\p B_1\cup \p B_2$ could be its limit from the interior or the exterior.
\end{theorem}
\pf If $\tau<0.5$, the gradient of $(u-H)$ is bounded independently of $\ep$ from \eqnref{gblowup2} and so does $\nabla q$ from Lemma \ref{lem:qbound}. Hence we can suppose that $\tau\geq 0.5$.
We prove the theorem using \eqnref{nor_bipolar}, \eqnref{tan_bipolar}, Lemma \ref{lem:bipolar} and Lemma \ref{lemma_sum}.
\qed
\smallskip

\begin{theorem}\label{main_thm_u_y}
For $0<k_1,k_2<1$, the solution $u$ to \eqnref{cond_eqn} can be expressed as
\beq\label{eqn:main_thm2}
u(\Bx) = c_{\mathbf{t}} \Im\{q(\Bx;\beta,-\tau_1,-\tau_2)\}+ H(\Bx)+ u_b(\Bx),\
c_{\mathbf{t}}=r_*^2 (\nabla H(\mathbf{p})\cdot\mathbf{t}),
\eeq
and there is a constant $C$ independent of $\ep$, $k_1$ and $k_2$
such that \beq\notag \|\nabla u_b\|_{L^\infty(\RR^2)}\leq C.\eeq
Here $|\nabla u_b|$ on $\p B_1\cup \p B_2$ could be its limit from the interior or the exterior.
\end{theorem}

The function $q$ is continuous in $\RR^2$ and harmonic except on $\p B_j$, and it decays to a constant at infinity, see Lemma \ref{qdecay}.
If the inclusions have the extreme property ($k_1=k_2=\infty$ or $k_1=k_2=0$), then $\beta=0$, $L(z;\beta)=-\log(1+z)$ and
$$
q(\Bx;\beta,|\tau_1|,|\tau_2|)=\xi+i\theta + b ,\quad \Bx\in \RR^2\setminus \overline{B_1\cup B_1},$$
where $\nabla b$ is uniformly bounded independently of $\epsilon$, for the proof see \eqnref{q011}. Hence we can consider $L$ as a generalized complex logarithm and (the real part of) $q$ as a generalization of $h$ in \eqnref{h:apo}. As it will be shown in section \ref{sub:Lerch}, $q$ and $\nabla q$ can be represented in terms of the so called Lerch transcendent function, of which numerical calculation has been intensively studied and implemented in commercial softwares.

The decomposition of $u$ in the main results is valid for the interior as well as the exterior of inclusions. Using these, we can derive not only the optimal bounds but also the singular term of the gradient of $u$ explicitly. Let us consider gradient of $q$ on the boundary of inclusions.
In section \ref{section:results}, we show that
$$\bigr|\nabla q(\Bx;\beta,|\tau_1|,|\tau_2|)\bigr|_{\p B_j}^+= \left(|\tau_1|+|\tau_2|+2\tau\right)\frac{\cosh\xi_j+\cos\theta}{2r^*\sqrt\ep} \left|\Re\{P\bigr(e^{-(\xi_j+i\theta)};\beta\bigr)\}\right|+O(1),$$
where
\beq \label{PL} P(z;\beta):=(-z)\pd{}{z}L(z;\beta).
\eeq
Especially at $\Bx_j$,{\it i.e.}, $\theta=0$, we have $\bigr|\nabla q(\Bx_j;\beta,|\tau_1|,|\tau_2|)\bigr|^+\approx \mbox{const.}{\ep^{-1/2}} \Re\{P\bigr(e^{-\xi_j};\beta\bigr)\}$. From the definition of $L$ which is an integral whose integrand contains $t^\beta$ term, $\Re\{P\bigr(e^{-\xi_j};\beta\bigr)\}$ is of order of $1/(\beta+1)$. This is in accordance with \eqnref{gblowup} and \eqnref{gblowup2}, see Remark \ref{remark1}.

While $\beta$ is zero for the extreme cases, it could be arbitrary large as well as small for highly conducting or almost insulating cases.
For example,
$\beta\approx{\epsilon}^{1/4}$ if $k_1,k_2\approx\epsilon^{-3/4}$ or $k_1,k_2\approx\epsilon^{3/4}$. Similarly, $\beta\approx\ep^{-1/4}$ if $k_1,k_2\approx\ep^{-1/4}$ or $k_1,k_2\approx\ep^{1/4}$. As it will be discussed in more detail later in the paper, one can derive the asymptotic of $P$ when $\beta$ is either large or small in comparison with $\ep$. Especially, for $\beta=O(\ep^{1/4})$,
\beq\notag\bigr|\nabla (u-u_\infty)\bigr|^+_{\p B_j}= \mbox{const.}\frac{1}{\ep^{1/4}}(\cosh\xi_j+\cos\theta)\ln[2(\cosh\xi_j+\cos\theta)]+O(1),\eeq
where $u_\infty$ is the solution of the perfectly conducting equation \eqnref{u_inf:eqn}.
In a practical computation of the electric field, \eqnref{u_inf:eqn} is often used instead of \eqnref{u:trans} when the inclusions are highly conducting. It is worth to emphasize that the error can be arbitrary large if so. More discussion is in section \ref{sub:nonuniform}.

The paper is organized as follows. In section 2, we review the bipolar coordinate system and derive a summation lemma which is essential to prove the main results. Section 3 is to derive the series expansion of $u$ in terms of bipolar coordinate system, and in section 4 we obtain the asymptotic of $\nabla u$.  We consider the conductivity problem defined in a bounded region in section 5 and illustrate the main results with numerical calculations in section 6.

\section{Preliminary}
\subsection{Bipolar coordinates}\label{subsection:bipolar}
Let us put \begin{equation}\label{def:a}\alpha:=\frac{|\mathbf{p}_1 -\mathbf{p}_2|}{2}=r_*\sqrt{\ep}+O(\ep).\end{equation}
In other words,
$$
\alpha = \frac{\sqrt{\ep (2 r_1 + \ep) (2 r_2 + \ep) (2 r_1 + 2 r_2 +
  \ep)}}{2 r_1 + 2 r_2 + 2\ep}.
$$

After rotation and shifting if necessary, it can be assumed that the centers of $\Bp_1$ and $\Bp_2$ are on the $x$-axis and  \begin{equation}\label{assump}\mathbf{p}_1=(-\alpha,0)\mbox{ and }\mathbf{p}_2=(\alpha,0).
\end{equation}
We assume so in what follows.

Each point $\mathbf{x}=(x,y)$ in the Cartesian coordinate system corresponds to  $(\xi,\theta)\in\RR\times (-\pi,\pi]$ in the bipolar coordinate system through the equations
\begin{equation} \label{bipolar}
x=\alpha\frac{ \sinh \xi }{\cosh \xi + \cos \theta} \quad \mbox{ and } \quad y=\alpha\frac{\sin \theta}{\cosh \xi + \cos \theta}
\end{equation}
with a positive number $\alpha$, see \cite{MF}. Setting $\alpha$ as \eqnref{def:a}, \eqnref{bipolar} means
\beq\label{def:bipolar}
\xi(\mathbf{x})=\ln({\Gamma_1}/{\Gamma_2})\mbox{ and }\theta(\mathbf{x}) \equiv \varphi_1-\varphi_2 \ (\mbox{mod } 2\pi)\eeq
for $$x+iy+\alpha=\Gamma_1 e^{i\varphi_1}\mbox{ and } \alpha-(x+iy)=\Gamma_2 e^{i\varphi_2}.$$
The magnitude $r$ of $\Bx$ satisfies
$$r(\xi, \theta)=\alpha\sqrt{(\cosh\xi -\cos\theta)/({\cosh\xi+\cos\theta})}.$$
Hence $|\Bx|\rightarrow +\infty$ if and only if $(\xi,\theta)\rightarrow(0, \pi)$, and if this is the case
\beq\label{rinf}\left|r(\xi,\theta)\sqrt{\cosh\xi+\cos\theta}\right|\leq 2\alpha.\eeq

From the definition, we can derive that the coordinate curve  $\{\xi=c\}$ and $\{\theta=c\}$ are, respectively, the zero-level set of
\begin{align}\ds\label{function:f}f_\xi(x,y)&=\left(x-\alpha\frac{\cosh c}{\sinh c}\right)^2 +y^2-\left(\frac{\alpha}{\sinh c}\right)^2,\\\ds
f_{\theta}(x,y)&=x^2 +\left(y+\alpha\frac{\cos c}{\sin c}\right)^2-\left(\frac{\alpha}{\sin c}\right)^2.\notag
\end{align}
Note that $\{|\theta|\leq\pi/2\}$ is the disk with the diameter $[-\alpha,\alpha]$, and $\{\pi/2\leq|\theta|\leq\pi-\ep^{1/4}\}$ is contained $\{|\Bx|\leq (2\alpha)/\sin (\ep^{\frac{1}{4}})\}$, whose radius is of magnitude $\ep^{1/4}$. Reversely, $\{|\Bx|\leq\ep^{1/4}\}\setminus(B_1\cup B_2)$ is contained in $\{|\theta|\leq \pi-\theta_\ep\}$ with $\theta_\ep\approx\ep^{1/4}$. That means the narrow region in between two inclusions cover almost all angle in the bipolar coordinate system. The graph in Figure \ref{fig:fig_nonuniform} illustrates coordinate curves of a bipolar coordinate system.
\begin{figure}[h!]
\begin{center}
\epsfig{figure=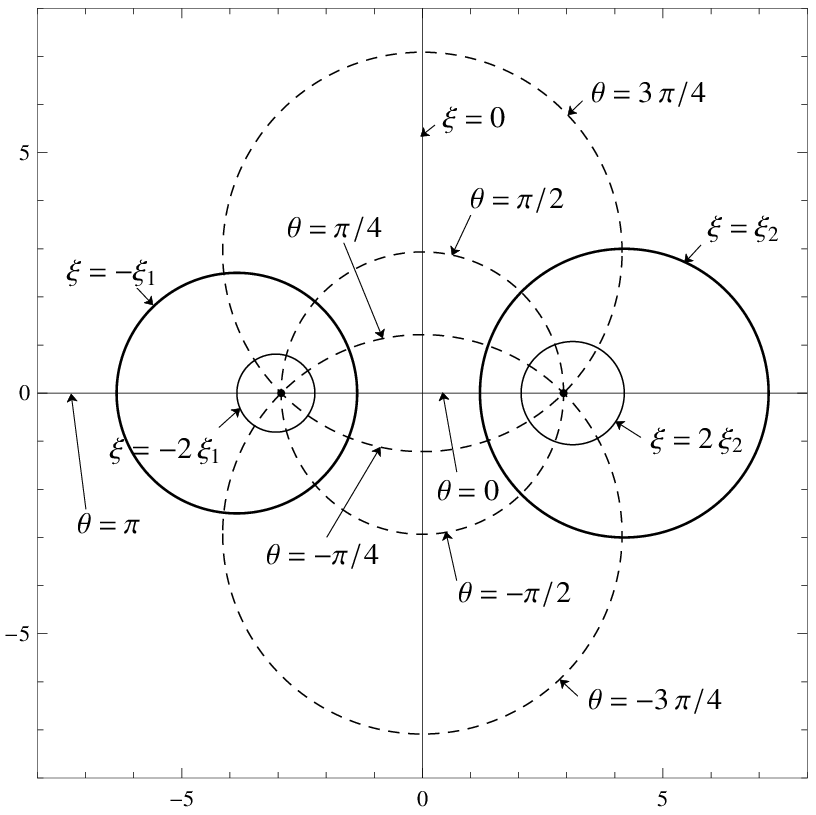,width=4.5cm}\hskip .5cm
\epsfig{figure=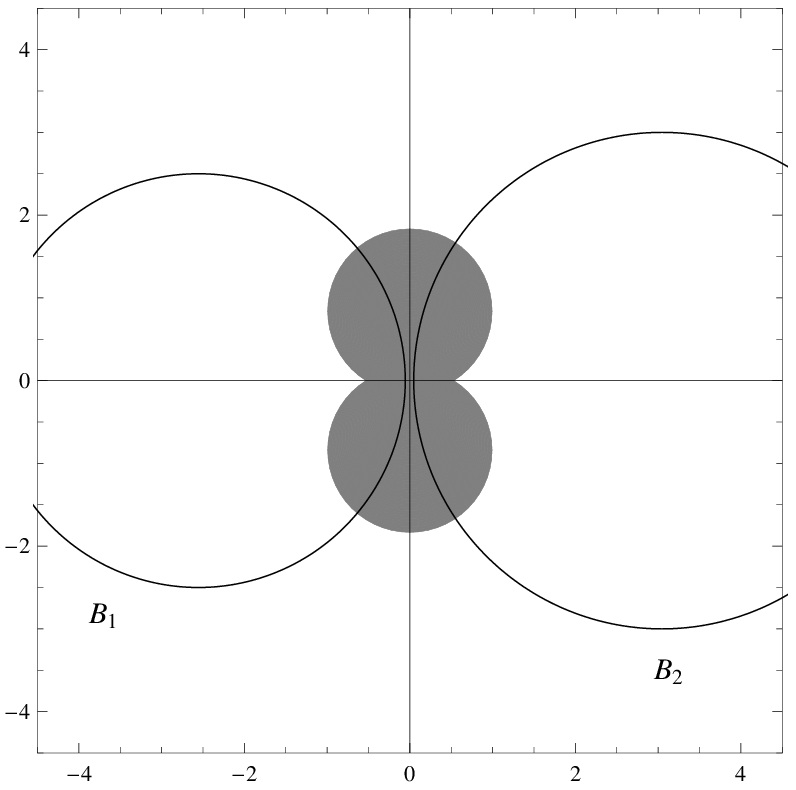,width=4.5cm}
\end{center}\label{fig:fig_nonuniform}
\vskip -.5cm\caption{Bipolar coordinate system for $r_1=2.5, r_2=3$. The distance $\ep$ between disks are 2 in the left and 0.1 in the right figure;
$(\xi_1,\xi_2)$ is $(0.99927,0.86602)$ and $(0.21021, 0.17557)$ in the left and right figure, respectively. As $\ep$ decreases, so do $\xi_1$ and $\xi_2$ in the order of $\sqrt\ep$, see \eqnref{eqnc1c2}.
In the left figure, solid lines are the coordinate curves of $\xi$-variable and the dotted ones are that of $\theta$. The shadow region in the right figure is $\{-\pi+\ep^{1/4}\leq \theta\leq\pi-\ep^{1/4}\}$. All points outside the shadow region have the bipolar coordinate $\theta$ close to $\pm\pi$.}
\end{figure}

Since $\p B_j\mbox{ is the coordinate curve }\{\xi=(-1)^j\xi_j\}$ for $\xi_j$ defined as \eqnref{def:xi12},
we have \beq\label{eqnc1c2} \xi_j=\sinh^{-1}\left(\frac{\alpha}{r_j}\right)\mbox{ and } \mathbf{c}_j=\alpha\bigr((-1)^j\coth\xi_j,0\bigr),\quad \mbox{for }j=1,2.\eeq
Hence $\xi_j={\alpha}/{r_j}+O(\ep\sqrt\ep)$, $j=1,2$, and
\begin{align}\label{axi1xi2} \frac{\alpha}{\xi_1+\xi_2}&=\frac{r_*^2}{2}+O({\epsilon}).\end{align}

 From \eqnref{function:f}, the outward unit normal $\nu$ to the circle $\xi=c$ for nonzero $c$ is
\beq\notag\nu_{\xi=c}=\frac{\nabla f_\xi}{|\nabla f_\xi|}=\mbox{sgn}(c)\left(\frac{-1-\cos\theta\cosh c }{\cosh c+\cos\theta}, \ \frac{-\sin\theta\sinh c}{\cosh c+\cos\theta}\right),\eeq
where sgn$(c)$ takes the value of 1 or -1 as $c$ is positive or negative, respectively. We define the tangential vector $T$ as the rotation of $\nu$ by $\pi/2$-radians.
One can see that
$$\nu_{\xi=c}=-\mbox{sgn}(c)\hat{\mathbf{e}}_\xi,\
T_{\xi=c}=-\mbox{sgn}(c)\hat{\mathbf{e}}_\theta,$$
where unit orthogonal basis vectors $\{\hat{\mathbf{e}}_{\xi},\hat{\mathbf{e}}_{\theta}\}$ are defined as
$$
\hat{\mathbf{e}}_{\xi}:= \frac{\p \mathbf{x}/ \p \xi}{|\p \mathbf{x}/ \p\xi|} \quad \mbox{and} \quad \hat{\mathbf{e}}_{\theta}:= \frac{\p \mathbf{x}/ \p \theta}{|\p \mathbf{x}/ \p\theta|}.
$$

It can be easily shown that the gradient of any scalar valued function $g$ is written in the following form:
\beq \label{grad_bipolar}
\nabla g = \frac{\cosh \xi+\cos \theta}{\alpha}\left( \frac{\p g}{\p\xi}\hat{\mathbf{e}}_{\xi}+ \frac{\p g}{\p\theta}\hat{\mathbf{e}}_{\theta}\right).
\eeq
Hence the normal- and tangential derivatives of a function $u$ in bipolar coordinates are \begin{align}\label{nor_bipolar}\pd{u}{\nu}\Bigr|_{\xi=c}&=\nabla u\cdot v_{\xi=c}=-\mbox{sgn}(c)\left(\frac{\cosh c+\cos\theta}{\alpha}\right)\pd{u}{\xi}\Bigr|_{\xi=c}\\
\label{tan_bipolar}\pd{u}{T}\Bigr|_{\xi=c}&=-\mbox{sgn}(c)\left(\frac{\cosh c+\cos\theta}{\alpha}\right)\pd{u}{\theta}\Bigr|_{\xi=c}.
\end{align}

The bipolar coordinate system is an orthogonal coordinate system and admits a general separation of variables solution to the harmonic function $f$ as follows:
\begin{align*}
f(\xi,\theta)&=a_0+b_0\xi+c_0\theta+\sum_{n=1}^\infty\bigr[ (a_n e^{n\xi}+b_n e^{-n\xi})\cos n\theta+\bigr (c_n e^{n\xi}+d_n e^{-n\xi})\sin n\theta\bigr],
\end{align*}
where $a_n$, $b_n$, $c_n$ and $d_n$ are constants.
Especially, the two linear functions $x$ and $y$ can be expanded as the following. For $\xi>0$, we have
\begin{align} \label{xy_zeta}
\frac{\sinh \xi -i\sin \theta}{\cosh \xi -\cos\theta} &=\frac{e^\zeta +e^{-\zeta}}{e^\zeta - e^{-\zeta}}=1+2\sum_{n=1}^\infty e^{-n\xi}(\cos n\theta-i\sin n\theta),
\end{align}
with $\zeta=\frac{\xi+i\theta}{2}$.
Plugging $(\theta+\pi)$ instead of $\theta$ and using \eqnref{bipolar},
\begin{align*}\ds
x&=\mbox{sgn}({\xi})\alpha\bigr[1+2\sum_{n=1}^\infty (-1)^n  e^{-n|\xi|}\cos n\theta\bigr],\\
\ds y&=-2\alpha\sum_{n=1}^\infty (-1)^n  e^{-n|\xi|}\sin n\theta.\end{align*}

\subsection{The Lerch transcendent function}\label{sub:Lerch}
The Lerch transcendent function $\Phi$ is defined as
\beq\label{Lerch}
\Phi(z,s,\beta):=\sum_{n=0}^{\infty} \frac{z^n}{(n+\beta)^s} \quad \mbox{ for } s\in \mathbb{C}, |z|<1 \mbox{ and } \beta \neq 0,-1,-2,... .
\eeq
The Lerch transcendent function has an integral representation
$$\Phi(z,1,\beta)=\int_0^1 \frac{t^{\beta-1}}{1-zt}\ dt$$
since we have
\begin{align}
\Phi(z,1,\beta) &= \sum_{n=0}^{\infty} z^n\int_0^1 t^{n+\beta-1} dt\nonumber=\int_0^1 t^{\beta-1} \sum_{n=0}^{\infty} \big(z t\big)^n dt.\notag
\end{align}
Here, interchanging the integral and series is possible due to uniform convergence of the series for $|z|<1$. Applying the change of variables,
\beq\Phi(z,1,\beta) =\int_0^\infty \frac{e^{-\beta t}}{1- z e^{-t}}\ dt.\label{Lerch_int}\eeq

One can easily show that
$$L(z;\beta)=-z\Phi(-z,1,\beta+1)$$
and, from \eqnref{PL},
\beq\label{PPhi}
P(z;\beta)=\frac{z}{1+z}-\beta z \Phi(-z,1,\beta+1).
\eeq
The function $P$ is crucial to understand the blow-up feature of $\nabla u$. To understand better, let us replace the summation in $P$ by an integral in terms of the function
\beq\notag
p_\theta(t):= \frac{1}{1+e^{-t+i\theta}}\quad \mbox{ defined for } t > 0,\ \theta\in(-\pi,\pi].
\eeq
Using \eqnref{Lerch_int}, we can express $P$ as
\beq\label{P1118}
\ds P(e^{-s+i\theta};\beta)=
-p_\theta(s)+ \beta\int_{0}^{\infty} e^{-\beta t}p_\theta(t+s) dt,\quad s>0.
\eeq
Applying the integration by parts, it becomes
\beq\label{integral_b}
\ds P(e^{-s+i\theta};\beta)=\int_0^\infty e^{-\beta t}{p_\theta\rq{}}(t+s)dt.
\eeq

\subsection{Properties of $P$}
Recall that
\begin{align}\label{ptheta:def}
p_\theta(t)=
\frac{e^t+\cos\theta-i\sin\theta}{2(\cosh t+\cos\theta)}=\frac{1}{2}+\frac{\sinh t-i\sin\theta}{2(\cosh t+\cos\theta)},
\end{align}
and the derivative of $p_\theta$ satisfies
\beq\label{pprime:0725} p_{\theta}'(t)=\frac{e^{-t+i\theta}}{(1+e^{-t+i\theta})^2}=\frac{1+\cosh t\cos\theta+i\sinh t \sin\theta}{2(\cosh t+\cos\theta)^2}.\eeq
Since
$\left|1+e^{-t+i\theta} \right|=\sqrt{2e^{-t}(\cosh t+\cos\theta)}$, we have
\begin{align}\label{0724:pbound}
\ds|p_\theta(t)|&\leq
\begin{cases}
\ds\frac{2}{\sqrt{\cosh s+\cos\theta}},\quad&\mbox{for }0<s\leq t<1\\[2mm]
\ds 2,&\mbox{for }t\geq 1,
\end{cases}\\
\label{pprime}
|p_\theta'(t)|&=\frac{1}{{2(\cosh t+\cos\theta)}},\quad t>0.\end{align}

\begin{lemma}\label{a_ep_bdd}Set $\beta, s, s_1,s_2>0$ and $s_1<s_2$. For all $\theta\in(-\pi,\pi]$,
\begin{align}\label{P1016}
\left |P(e^{-s+i\theta};\beta) \right| \leq \frac{1}{2\beta (\cosh s+\cos\theta)},\\
\label{P1016_2}
\left |P(e^{-s+i\theta};\beta) \right| \leq 4+\frac{4}{\sqrt{\cosh s+\cos\theta}},\\
\label{compare:0729}\left|P(e^{-s_2+i\theta};\beta)-P(e^{-s_1+i\theta};\beta)\right|\leq\frac{s_2-s_1}{\cosh s_1+\cos\theta}.\end{align}
\end{lemma}
\pf
If $s>1$, then we have $|P(\theta;s,\beta)|\leq 4$ from \eqnref{0724:pbound}. Set $s<1$, then
$$|p_\theta(t)|\leq \frac{2}{\sqrt{\cosh s+\cos\theta}}\qquad\qquad \mbox{for }s\leq t<1.$$
Applying \eqnref{P1118}, it becomes
\begin{align*}
\left |P(e^{-s+i\theta};\beta) \right| &\leq \frac{2}{\sqrt{\cosh s+\cos\theta}}\left(1+\beta \int_0^{1-s} e^{-\beta t}\ dt\right)+\beta\int_{1-s}^\infty e^{-\beta t}2dt\\&\leq \frac{4}{\sqrt{\cosh s+\cos\theta}}+2.
\end{align*}
It proves \eqnref{P1016_2}.
Moreover, from \eqnref{integral_b} and \eqnref{pprime}, we prove \eqnref{P1016}.

Let $t\geq0$. From the mean value property,
\begin{align}\notag
p_\theta (t+s_2) - p_\theta(t+s_1)&=(s_2-s_1)  p_\theta'(c)\quad\mbox{for some }c\in (t+s_1, t+s_2).
\end{align}
Applying \eqnref{pprime},
\beq
\left|p_\theta(t+s_2) - p_\theta(t+s_1)\right|\leq \frac{s_2-s_1}{2(\cosh s_1+\cos\theta)}.\notag
\eeq
Therefore we have
\beq\notag\left|p_\theta(s_2)-p_\theta(s_1)\right|\leq \frac{s_2-s_1}{2(\cosh s_1+\cos\theta)},\eeq
\begin{align*}
\left|\beta\int_0^\infty e^{-\beta t}p_\theta(t+s_2)\ dt
- \beta \int_0^\infty e^{-\beta t} p_\theta(t+s_1) dt \right|
&\leq \beta \int_0^\infty e^{-\beta t} \frac{s_2-s_1}{2(\cosh s_1+\cos\theta)}dt.
\end{align*}
This proves \eqnref{compare:0729}.
\qed

\begin{lemma}\label{at0}
Let $\beta,\xi>0$. For all $\theta\in(-\pi,\pi]$, we have
\begin{align}\label{P1016_3}\Bigr|(\cosh \xi+\cos\theta)\  P(e^{-\xi+i\theta};\beta)\Bigr|\leq\frac{e^\xi}{2(\beta+1)},\\
\frac{1}{4(\beta+1)}\leq\Re \{P(e^{-\xi};\beta)\}\leq\frac{1}{\beta+1},\quad\mbox{at }\theta=0.\end{align}
\end{lemma}
\pf
When $\theta=0$, from \eqnref{pprime:0725} and \eqnref{integral_b},
\begin{align*}
\frac{1}{4}\int_0^\infty e^{-(\beta+1)t}dt\leq  \Re \{P(e^{-\xi};\beta)\}\leq\int_0^\infty e^{-(\beta+1)t}dt=\frac{1}{\beta+1}.
\end{align*}
From  \eqnref{P1118} and \eqnref{ptheta:def},
\beq\nonumber\Re\{P(e^{-\xi+i\theta};\beta)\}=-\frac{\sinh\xi}{2(\cosh \xi+\cos\theta)}+\beta\int_0^\infty e^{-\beta t}\frac{\sinh (t+\xi)}{2(\cosh (t+\xi)+\cos\theta)} dt.\eeq
Hence
\begin{align*}
0&<(\cosh \xi+\cos\theta)\ \Re \{P(e^{-\xi+i\theta};\beta)\}+\frac{\sinh\xi}{2}= \beta\int_0^\infty e^{-\beta t}\frac{(\cosh \xi+\cos\theta)\sinh (t+\xi)}{2(\cosh (t+\xi)+\cos\theta)} dt\\
&\leq  \beta\int_{0}^{\infty} e^{-\beta t}\frac{(\cosh\xi+ 1)\sinh (t+\xi)}{2(\cosh( t+\xi)+1)}\ dt=(\cosh \xi+1)\ \Re \{P(e^{-\xi};\beta)\}+\frac{\sinh\xi}{2}.
\end{align*}
\qed

\subsection{Summation formula}\label{section:summation}
In this section we prove a lemma which is essential to prove the main results in this paper.

\begin{lemma}\label{lemma_sum0}
Fix a $a_0>0$. For $a<a_0$, $0<\tau<1$ and $\theta\in(-\pi,\pi]$, we have
\begin{align*}
 \ds&\left|a_0\sum_{m=1}^{\infty}\tau^{m-1} \frac{e^{-ma_0+a+i\theta}}{(1+ e^{-ma_0+a+i\theta})^2}-P(e^{-(a_0-a)+i\theta}; -(\ln\tau)/a_0)\right|\leq\frac{4a_0}{\cosh (a_0-a)+\cos \theta}.
\end{align*}
\end{lemma}
\pf
Fix a $\theta$. Thanks to \eqnref{pprime:0725},
\begin{align}\label{sum:0724}
\ds&a_0\sum_{m=1}^{\infty} \tau^{m-1} \frac{e^{-ma_0+a+i\theta}}{(1+ e^{-ma_0+a+i\theta})^2}=a_0\sum_{m=0}^{\infty} \tau^m p_\theta' (ma_0+s_0),\quad s_0=a_0-a>0.
\end{align}

To rewrite the right-hand side of \eqnref{sum:0724}, we consider $f(t):=\tau^tp_\theta'(a_0t+s_0)$.
From \eqnref{pprime} and the fact that $\tau\in(0,1)$,
\beq\label{gmaxat0}
\ds\left|\tau^t p_\theta'(a_0t+s_0)\right|\leq\frac{1}{2(\cosh s_0+\cos\theta)},\quad t\geq 0,\eeq
and the derivative of $f$ is as follows:
\beq\label{pprime2}\frac{d}{dt}\Bigr(\tau^t p_\theta'(a_0t+s_0)\Bigr)=\tau^t\Bigr((\ln \tau)  p_\theta' (a_0t+s_0) + s p_\theta''(a_0t+s_0)\Bigr).\eeq
 Note that, from \eqnref{pprime:0725}, the real and imaginary part of $e^{2(a_0t+s_0)} \Bigr((\ln \tau)  p_\theta'(a_0t+s_0) +a_0 p_\theta''(a_0t+s_0)\Bigr)$
 are quadratic polynomials of $e^{a_0t+s_0}$. Hence  the real and imaginary part of $\frac{d}{dt}\Bigr(\tau^t p_\theta'(a_0t+s_0)\Bigr)$ changes the sign at most 4 times on $(0,\infty)$.
Therefore, from \eqnref{pprime},  $$\left|a_0\sum_{m=0}^{\infty} \tau^m p_\theta' (ma_0+s_0)-a_0\int_0^\infty \tau^t p_\theta'(a_0t+s_0) dt\right|\leq 8\sup_{t\geq0}|p_\theta'(t)|\leq\frac{4a_0}{{\cosh s_0+\cos\theta}}.$$
By a change of variables, we get
\begin{align}
\ds a_0\int_0^\infty \tau^t p_\theta'(a_0t+s_0) dt&=\int_{0}^{\infty} e^{-b t} p_\theta'(t+s_0) dt=-p_\theta(s_0)+b\int_0^\infty e^{-bt}p_\theta(t+s_0)\ dt\notag
\end{align}
with $b=\frac{-\log \tau}{a_0}$. This proves the lemma.
\qed

\section{Expansion of $u$ in terms of the bipolar system}
From \eqnref{usur}, we only need to consider linear functions for $H$.
We can also assume $B_1$ and $B_2$ satisfy \eqnref{assump} without loss of generality. Define the bipolar system $(\xi,\theta)$ and $\xi_j$, $j=1,2$, as in section \ref{subsection:bipolar}.
So $\xi=-\xi_1$ and $\xi=\xi_2$ represents $\p B_1$ and $\p B_2$, respectively.
 Set $$\xi_M:=\max(\xi_1,\xi_2)\mbox{ and }\xi_s:=\min(\xi_1,\xi_2),$$ then
\beq\label{xirange}
\begin{cases}-\xi-2\xi_1, \xi-2\xi_2\in (-2\xi_M-\xi_s, -\xi_s),\ &\Bx\in\RR^2\setminus\overline{B_1\cup B_2}\\
\xi\in(-\infty, -\xi_1),&\Bx\in B_1\\
\xi\in (\xi_2,\infty),&\Bx\in B_2.
\end{cases}
\eeq
\begin{lemma}\label{lem:bipolar}
Define a complex valued function
\begin{equation} \label{sol_cond}
 \ds U(x,y):=C+\begin{cases}
 \ds \sum_{n=1}^{\infty} \left(A_n e^{n (\xi+i\theta)} + B_n e^{-n( \xi+i\theta)}\right) ,\quad &x\in\mathbb{R}^2 \setminus (B_1 \cup B_2)\\
 \ds \sum_{n=1}^\infty \left(A_n e^{n(\xi+i\theta)}+B_ne^{n(2\xi_1+\xi-i\theta)}\right),& x\in B_1\\
 \ds  \sum_{n=1}^\infty\left(A_ne^{n(2\xi_2-\xi+i\theta)}+B_n e^{-n( \xi+i\theta)}\right),& x\in B_2
 \end{cases}
\end{equation}
where $C=-\sum_{n=1}^{\infty} (A_n + B_n )\cos n\pi$
and
\begin{align*}
\ds A_n&=\frac{2\alpha(-1)^n}{\tau^{-1} e^{2n(\xi_1+\xi_2)}-1}\left(-\frac{1}{\tau_1}e^{2n\xi_1}-1\right),\
\ds B_n=\frac{2\alpha(-1)^n}{\tau^{-1} e^{2n(\xi_1+\xi_2)}-1}\left(1+\frac{1}{\tau_2}e^{2n\xi_2}\right).
\end{align*}
Then $(x+\Re\{U\})$ is the solution to \eqnref{cond_eqn} for $H(x,y)=x$, and $(y+\Im\{U\})$ is the solution for $H(x,y)=y$ with $1/k_j$ in the place of $k_j$, $j=1,2$.
\end{lemma}
We prove the lemma after the following one.

\begin{lemma}\label{lemma_sum} Suppose $\tau\geq 0.5$. There is a constant $C$ depending only on $r_1,r_2$ such that
$$\left|\pd{U}{\xi}(\Bx)-r_*^2\pd{}{\xi}q(\Bx;\beta,\tau_1,\tau_2)\right|\leq \frac{C}{\cosh\xi_s+\cos\theta}, \quad\Bx\in\RR^2,$$ where $q$ is defined as \eqnref{a_ep}.
We have the same equation for the partial derivative in $\theta$ variable.
\end{lemma}
\pf
Note that $A_n$ and $B_n$ can be represented as
\begin{align}A_n &= 2\alpha(-1)^n\sum_{m=1}^\infty \tau^m\left(-\frac{1}{\tau_1}\ e^{2n\xi_1}-1\right)e^{-2nm(\xi_1+\xi_2)},\notag\\
B_n&=2\alpha(-1)^n\sum_{m=1}^\infty \tau^m\left(1+\frac{1}{\tau_2}\ e^{2n\xi_2}\right)e^{-2nm(\xi_1+\xi_2)}.\notag
\end{align}

Firstly, set $-\xi_1\leq\xi\leq\xi_2$. By interchanging the order of summation, which is possible due to the absolute convergence, we have
\begin{align}
\ds &\sum_{n=1}^{\infty} n\left(A_n e^{n(\xi+i\theta)}- B_n e^{-n(\xi+i\theta)}\right)\nonumber\\& = -2\alpha  \sum_{m=1}^{\infty} \tau^m \Bigg[\ \frac{1}{\tau_1}  \ds\sum_{n=1}^{\infty} n  \left( -e^{-2m(\xi_1+\xi_2)+2\xi_1+\xi+i \theta} \right)^n   + \sum_{n=1}^{\infty} n \left( -e^{-2m(\xi_1+\xi_2)+\xi+i\theta}\right)^n \nonumber \\
& \quad\quad\ds+ \frac{1}{\tau_2}  \sum_{n=1}^{\infty} n  \left( -e^{-2m(\xi_1+\xi_2)+2\xi_2-\xi-i \theta} \right)^n + \sum_{n=1}^{\infty} n \left( -e^{-2m(\xi_1+\xi_2)-\xi-i\theta}\right)^n \Bigg].\label{sum:0808}
\end{align}
Using the following formula
$
\sum_{n=1}^{\infty} n z^n= \frac{z}{(1-z)^2},\mbox{ for }|z|<1,\ z\in\CC,$ we have
\begin{align}
\ds &\sum_{n=1}^{\infty} n\left(A_n e^{n(\xi+i\theta)}- B_n e^{-n(\xi+i\theta)}\right)\nonumber\\
&= 2\alpha\tau \sum_{m=1}^{\infty} \tau^{m-1} \Bigg[ \frac{1}{\tau_1} \frac{e^{-2m(\xi_1+\xi_2)+2\xi_1+\xi+i \theta}}{(1+e^{-2m(\xi_1+\xi_2)+2\xi_1+\xi+i  \theta})^2}+ \frac{e^{-2m(\xi_1+\xi_2)+\xi+i  \theta}}{(1+e^{-2m(\xi_1+\xi_2)+\xi+i  \theta})^2}\nonumber \\\label{sum:0808_2}
& \quad \quad  +\frac{1}{\tau_2} \frac{e^{-2m(\xi_1+\xi_2)+2\xi_2-\xi-i \theta}}{(1+e^{-2m(\xi_1+\xi_2)+2\xi_2-\xi-i  \theta})^2}
+\frac{e^{-2m(\xi_1+\xi_2)-\xi-i \theta}}{(1+e^{-2m(\xi_1+\xi_2)-\xi-i  \theta})^2}  \Bigg].\end{align}
We apply Lemma \ref{lemma_sum0} to this series with $a=2(\xi_1+\xi_2)$.
From \eqnref{sum:0808}, \eqnref{sum:0808_2}, Lemma \ref{lemma_sum0}, and Lemma \ref{a_ep_bdd},
\begin{align*}
&\sum_{n=1}^{\infty} n\left(A_n e^{n(\xi+i  \theta)}- B_n e^{-n(\xi+i  \theta)}\right)\\
&=\frac{{r_*^2}}{2}\Bigg[\tau_2 P(e^{-2 \xi_2+\xi+i\theta};b)+\tau P(e^{-2(\xi_1+\xi_2)+\xi+i\theta} ; b)+\tau_1 P(e^{-2\xi_1-\xi-i\theta};b)
\\&\qquad\qquad+ \tau P(e^{-2(\xi_1+\xi_2)-\xi-i\theta};b)\Bigg]+\frac{r_*^2(\tau_1+\tau_2+2\tau_1\tau_2)}{2}\frac{O(\sqrt\ep)}{\cosh \xi_s+\cos\theta},
\end{align*}
where $$b=\frac{-\ln\tau}{2(\xi_1+\xi_2)}.$$

Note that $b=\beta + O(\sqrt{\epsilon})$ where $\beta$ is defined as \eqnref{parameter_beta}.
Since $b,\beta>0$, applying the mean value property for a fixed $t>0$, we have
\beq\label{exp_mv}be^{-b t} =\beta e^{-\beta t} + (b-\beta)(1-st)e^{-st},\quad\mbox{for a }s \in(\min(b,\beta),\ \max(b,\beta)).\eeq
Hence
\beq\label{beb}
\ds be^{-b t} -\beta e^{-\beta t} =
\begin{cases}(b-\beta)O(1),\quad& \mbox{for }t>0,\\
(b-\beta)O(t)e^{-\min(b,\beta)t},\quad&\mbox{for }t\geq1.
\end{cases}
\eeq
Therefore, using \eqnref{0724:pbound} as well,
\begin{align*}
&\left|b \int_0^\infty e^{-b t} p_\theta(t+\xi_s) dt-\beta \int_0^\infty e^{-\beta t} p_\theta(t+\xi_s) dt\right|\\
&\leq|b-\beta|\int_0^1O(1)\frac{1}{\sqrt{\cosh \xi_s+\cos\theta}}\ dt+|b-\beta|\int_1^\infty e^{-\min{(b,\beta)}t} O(t) dt\\
&\ds\leq C\sqrt{\epsilon}\frac{1}{\cosh \xi_s+\cos\theta} +O(\sqrt\ep).
\end{align*}
Here the remainder term $O(\sqrt{\epsilon})$ is uniform with respect to $k_1,k_2,{\xi}$ and $\theta$. Therefore we can replace $b$ by $\beta$.
Using \eqnref{axi1xi2}, we prove the lemma for $-\xi_1\leq\xi\leq\xi_2$.

Similarly, for $ \xi \leq -\xi_1$, we can derive that
\begin{align*}
&\ds\sum_{n=1}^\infty n\left(A_n e^{n(\xi+i\theta)}\pm B_ne^{2n\xi_1}e^{n(\xi-i\theta)}\right)\\\ds&=\frac{{r_*^2}}{2}\Bigg[(\tau_2+\tau)P(e^{\xi+i\theta-2\xi_2};\beta)\mp(\tau_1+\tau)P(e^{\xi-i\theta};\beta)\Bigg]+\frac{O(\sqrt\ep)}{\cosh \xi_s+\cos\theta},\end{align*}
and, for $\xi\geq\xi_2$,
\begin{align*}
&\ds\sum_{n=1}^\infty n \left(\mp A_ne^{2\xi_2} e^{-n(\xi-i\theta)}-B_n e^{-n( \xi+i\theta)}\right)\\\ds&=\frac{{r_*^2}}{2}\Bigg[\mp(\tau_2+\tau)P(e^{-\xi+i\theta};\beta)+(\tau_1+\tau)P(e^{-\xi-i\theta-2\xi_1};\beta)\Bigg]+\frac{O(\sqrt\ep)}{\cosh \xi_s+\cos\theta}.
\end{align*}
This proves the lemma for  $ \xi \leq -\xi_1$ and for $\xi\geq\xi_2$.
\qed

\noindent{\large{\textbf{Proof of Lemma \ref{lem:bipolar}}}}
It can be easily shown that the function $u$ defined as \eqnref{sol_cond} satisfies the transmission condition of \eqnref{u:trans}.
Hence it is only need to show that there is a positive constant $M$ such that
\begin{equation}
\limsup_{|\mathbf{x}|\rightarrow\infty} |\mathbf{x}||(u-H)(\mathbf{x})|\leq M.
\end{equation}
From \eqnref{rinf}, it is enough to show that
$$\limsup_{(\xi,\theta)\rightarrow (0,\pi)}\frac{|(u-H)(\mathbf{x})|}{\sqrt{\cosh\xi+\cos\theta}}\leq M$$
when the distance $\ep$ is fixed.

Plugging $\xi=0$ and $\theta=\pi$ into \eqnref{sum:0808}, it becomes
\begin{align*}
\sum_{n=1}^{\infty} \bigr[(A_n + B_n )\cos n\pi\bigr]
&= 2\alpha\tau \sum_{m=1}^{\infty} \tau^{m-1} \Bigg[\frac{1}{\tau_1} \frac{-e^{-2m(\xi_1+\xi_2)+2\xi_2}}{1-e^{-2m(\xi_1+\xi_2)+2\xi_2}}\nonumber
+ \frac{-e^{-2m(\xi_1+\xi_2)}}{1-e^{-2m(\xi_1+\xi_2)}} \nonumber \\
& \qquad\qquad\quad \quad -\frac{1}{\tau_2} \frac{-e^{-2m(\xi_1+\xi_2)+2\xi_1}}{1-e^{-2m(\xi_1+\xi_2)+2\xi_1}}
-\frac{-e^{-2m(\xi_1+\xi_2)}}{1-e^{-2m(\xi_1+\xi_2)}}  \Bigg].\end{align*}
Note that for $\tilde\xi=0, 2\xi_1, 2\xi_2$ and $|\xi|\leq\xi_s$. Hence we have
\begin{align}
\left|\frac{e^{-2m(\xi_1+\xi_2)+\tilde\xi+\xi+i\theta}}{1+e^{-2m(\xi_1+\xi_2)+\tilde\xi+\xi+i\theta}}
-\frac{-e^{-2m(\xi_1+\xi_2)+\tilde\xi}}{1-e^{-2m(\xi_1+\xi_2)+\tilde\xi}}\right|\notag
&=\left|\frac{(e^{\xi+i\theta}+1)e^{-2m(\xi_1+\xi_2)+\tilde\xi}}{(1+e^{-2m(\xi_1+\xi_2)+\tilde\xi+\xi+i\theta})
(1-e^{-2m(\xi_1+\xi_2)+\tilde\xi})}\right|\notag\\\notag
&\leq\frac{e^{-2m(\xi_1+\xi_2)+\tilde\xi+\xi_s}}{(1-e^{-2m(\xi_1+\xi_2)+\tilde\xi+\xi_s})^2}|e^{\xi+i\theta}+1|.
\end{align}

Let us define a constant
\begin{align*}
S&:= 2\alpha\tau \sum_{m=1}^{\infty} \tau^{m-1} \Bigg[ \left(\frac{1}{\tau_1}\right) \frac{e^{-2m(\xi_1+\xi_2)+2\xi_2+\xi_s}}{(1-e^{-2m(\xi_1+\xi_2)+2\xi_2+\xi_s})^2}\nonumber
+ \frac{e^{-2m(\xi_1+\xi_2)+\xi_s}}{(1-e^{-2m(\xi_1+\xi_2)+\xi_s})^2}\nonumber \\
& \qquad\qquad\quad \quad  +\left(\frac{1}{\tau_2}\right) \frac{e^{-2m(\xi_1+\xi_2)+2\xi_1+\xi_s}}{(1-e^{-2m(\xi_1+\xi_2)+2\xi_1+\xi_s})^2}
+\frac{e^{-2m(\xi_1+\xi_2)+\xi_s}}{(1-e^{2m(\xi_1+\xi_2)+\xi_s})^2}  \Bigg].\end{align*}
Applying the same procedure as in the proof of Lemma \ref{lemma_sum} ,
we have
\begin{align*} |(u-H)(\Bx)|&=\left|\sum_{n=1}^\infty\bigr[(A_n e^{n \xi} + B_n e^{-n \xi})\cos n\theta- (A_n + B_n )\cos n\pi\bigr]\right|\\&\leq S|e^{\xi+i\theta}+1|.
\end{align*}
Hence
$\left|(u-H)(\mathbf{x})\right|\leq C|\xi+i(\theta-\pi)|,\mbox{ as }(\xi,\theta)\rightarrow (0,\pi),$
with a constant $C$ independent of $\xi$ and $\theta$.
\qed

\begin{lemma}\label{qdecay}
The singular function $q$ in \eqnref{a_ep} satisfies $$q(\Bx;\beta,\tau_1,\tau_2)-C_q = O(|\Bx|^{-1}),\quad \mbox{as } |\Bx| \rightarrow \infty,$$
Where $C_q$ is a constant defined as
$C_q=\frac{1}{2}\Bigr\{ (\tau_1+\tau)L(-e^{-2\xi_1};\beta)- (\tau_2+\tau)L(-e^{-2\xi_2};\beta)\Bigr\}.$
\end{lemma}
\pf
From the definition of $L$, we have
\begin{align*}
\ds\left|L(e^{-(\xi+i \theta)-2\xi_1};\beta)-L(e^{-2\xi_1};\beta)\right|
\ds&\leq \int_0^1 t^\beta\left|\frac{e^{-(\xi+i\theta)-2\xi_1}}{1+t e^{-(\xi+i\theta)-2\xi_1}}-
\frac{-e^{-2\xi_1}}{1-t e^{-2\xi_1}}\right|dt\\
\ds&=\int_0^1 t^\beta\left|\frac{(e^{-(\xi+i\theta)}+1)te^{-2\xi_1}}{(1+t e^{-(\xi+i\theta)-2\xi_1})(1-t e^{-2\xi_1})}\right|dt\\
\ds&=\left| e^{-(\xi+i\theta)}+1 \right|\int_0^1 t^\beta\left|\frac{te^{-2\xi_1}}{(1-t e^{-2\xi_1})^2}\right|dt\\
\ds&\leq C |\xi+i(\theta-\pi)|, \quad \mbox{as }(\xi,\theta)\rightarrow (0,\pi),
\end{align*}
where a constant $C$ is independent of $\xi$ and $\theta$.
Therefore, we have
$$
L(e^{-(\xi+i \theta)-2\xi_1};\beta)-L(e^{-2\xi_1};\beta)=O(|\Bx|^{-1}),\quad \mbox{as } |\Bx|\rightarrow \infty.
$$
Similarly, we prove the deacay property of $
L(e^{(\xi+i \theta)-2\xi_2};\beta)-L(e^{-2\xi_2};\beta)$.
\qed

\section{Asymptotic Estimates for $\nabla u$}\label{section:results}
\subsection{Estimates of the gradient of the singular function $q$}\label{sub:q}
The main term of $u$ in both Theorem \ref{main_thm_u_x} and Theorem \ref{main_thm_u_y} is expressed in terms of $q(\Bx;\beta,\tau_1,\tau_2)$ with positive $\tau_1$ and $\tau_2$. We assume that $\tau_1$ and $\tau_2$ are positive in this section.

For notational sake, the normal- and the tangential derivatives at $(c,\theta)$ mean \eqnref{nor_bipolar} and \eqnref{tan_bipolar}, respectively, on the coordinate level curve $\{\xi=c\}$.
From \eqnref{a_ep} and \eqnref{PL}, the derivative of $q(\Bx;\beta,\tau_1,\tau_2)$ satisfies
\begin{align}
&\pd{q}{\xi}= \frac{1}{2}
\begin{cases}
\ds (\tau_1+\tau) P\bigr(e^{-(\xi+i\theta)-2\xi_1};\beta\bigr)+(\tau_2+\tau)P\bigr(e^{(\xi+i\theta)-2\xi_2};\beta\bigr) \  &\mbox{in }\mathbb{R}^2 \setminus \overline{(B_1 \cup B_2)},\\[1mm]
 \ds  -(\tau_1+\tau)P\big(e^{\xi-i\theta};\beta\big)+(\tau_2+\tau)P\big(e^{\xi+i\theta-2\xi_2};\beta\big)\quad &\mbox{in }B_1,\\[1mm]
 \ds   (\tau_1+\tau)P\big(e^{-\xi-i\theta-2\xi_1};\beta\big)-(\tau_2+\tau)P\big(e^{-\xi+i\theta};\beta\big)\quad &\mbox{in }B_2,\\[2mm]
  \end{cases} \label{dqxi}\\
&\pd{q}{\theta}\notag=  \frac{i}{2}
\begin{cases}
\ds (\tau_1+\tau) P\bigr(e^{-(\xi+i\theta)-2\xi_1};\beta\bigr)+(\tau_2+\tau)P\bigr(e^{(\xi+i\theta)-2\xi_2};\beta\bigr) \  &\mbox{in }\mathbb{R}^2 \setminus \overline{(B_1 \cup B_2)},\\[1mm]
 \ds  (\tau_1+\tau)P\big(e^{\xi-i\theta};\beta\big)+(\tau_2+\tau)P\big(e^{\xi+i\theta-2\xi_2};\beta\big)\quad &\mbox{in }B_1,\\[1mm]
 \ds   (\tau_1+\tau)P\big(e^{-\xi-i\theta-2\xi_1};\beta\big)+(\tau_2+\tau)P\big(e^{-\xi+i\theta};\beta\big)\quad &\mbox{in }B_2,
  \end{cases}
\end{align}
where $P$ is defined as \eqnref{PL}. From \eqnref{nor_bipolar}, \eqnref{tan_bipolar}, \eqnref{P1016_2} and \eqnref{axi1xi2}
\begin{align}\label{qnu1}
(\nabla q\cdot \hat{\mathbf{e}}_\xi)\bigr|_{\xi=c}&=\frac{\cosh c+\cos\theta}{r_*\sqrt\ep}\pd{q}{\xi}\Bigr|_{\xi=c}+O(1),\\
\label{qtan1}
(\nabla q\cdot \hat{\mathbf{e}}_\theta)\bigr|_{\xi=c}&=\frac{\cosh c+\cos\theta}{r_*\sqrt\ep}\pd{q}{\theta}\Bigr|_{\xi=c}+O(1).\end{align}

\begin{lemma}\label{lem:qbound}
There is a constant $C$ independent of $\ep$, $k_1$, and $k_2$ such that
$$\|\nabla q\|_{L^\infty(\RR^2)}\leq C \mbox{ if }\tau<\frac{1}{2}.$$
\end{lemma}
\pf
Applying \eqnref{xirange} to \eqnref{P1016_2} for $\Bx\in B_1\cup B_2$ and \eqnref{P1016_3} for $\Bx\in\RR^2\setminus\overline{B_1\cup B_2}$, we can prove than there is a $C$ independent of $\epsilon$, $k_1$ and $k_2$ such that
\beq\label{qnubound}\left|(\nabla q\cdot \hat{\mathbf{e}}_\xi)\right|, \left|(\nabla q\cdot \hat{\mathbf{e}}_\theta)\right| \leq C\frac{1}{-\ln\tau},\eeq
for $c\neq -\xi_1,\xi_2$. At $c=(-1)^j\xi_j$, $j=1,2$, \eqnref{qnubound} is satisfied for either outward or inward derivatives.
\qed

\begin{prop}\label{cornor}
Suppose $\tau_1,\tau_2>0$.
The tangential derivative of $\Re\{q(\Bx;\beta,\tau_1,\tau_2)\}$ is remain bounded regardless of $\ep$ for all $\Bx\in\RR^2$. The normal derivative of $\Re\{q(\Bx;\beta,\tau_1,\tau_2)\}$ is bounded in $B_1\cup B_2$. For $\Bx\notin\overline{B_1\cup B_2}$,
\begin{align}&\pd{\Re\{q\}}{\nu}\bigg|_{\xi=c}=-\mbox{sgn}(c)\left(\tau_1+\tau_2+2\tau\right)\frac{\cosh\xi_s+\cos\theta}{2r_*\sqrt\ep}\Re\left\{ P\bigr(e^{-(\xi_s+i\theta)};\beta\bigr)\right\}+O(1).\label{cornoreqn}
\end{align}
\end{prop}
\pf
From the definition of $P$,
\beq\label{Pconj}
P(\overline{z};\beta)=\overline{P(z;\beta)},\quad\mbox{for all }z\in\CC, |z|<	1.
\eeq
Hence \beq\label{ReImP}
\Re\{P(\overline{z})\}=\Re\{P(z)\}\mbox{ and }\Im\{P(\overline{z})\}=-\Im\{P(z)\}.\eeq
From \eqnref{dqxi}, \eqnref{ReImP}, \eqnref{compare:0729} and \eqnref{qnu1}, we have
\begin{align*}
&\ds\pd{\Re\{q\}}{\nu}\bigg|_{\xi=c}=-\mbox{sgn}(c)
\begin{cases}\ds\frac{\cosh \xi_s+\cos\theta}{2r_*\sqrt\ep}\left(\tau_1+\tau_2+2\tau\right)\Re\left\{ P\bigr(e^{-\xi_s+i\theta};\beta\bigr)\right\}+O(1),\quad&\Bx\in\RR^2\setminus\overline{B_1\cup B_2}\\[2mm]
\ds\frac{\cosh \xi+\cos\theta}{2r_*\sqrt\ep}\left(\tau_2-\tau_1\right)\Re\left\{ P\bigr(e^{\xi+i\theta};\beta\bigr)\right\}+O(1),\quad&\Bx\in B_1\\[2mm]
\ds\frac{\cosh \xi+\cos\theta}{2r_*\sqrt\ep}\left(\tau_1-\tau_2\right)\Re\left\{ P\bigr(e^{-\xi+i\theta};\beta\bigr)\right\}+O(1),\quad&\Bx\in B_2,
\end{cases}
\end{align*}
\begin{align*}
&\ds\pd{\Re\{q\}}{T}\bigg|_{\xi=c}=-\mbox{sgn}(c)\frac{\cosh c+\cos\theta}{2r_*\sqrt\ep}\left(\tau_1-\tau_2\right)\Re\left\{ P\bigr(e^{-\max(\xi_s,|c|)-i\theta};\beta\bigr)\right\}+O,\quad\Bx\in\RR^2.
\end{align*}

Let us prove the boundedness part in the corollary.
From \eqnref{P1016}, we have
\beq\notag
\left|\pd{}{\nu}\Re\{q(\Bx;\beta,\tau_1,\tau_2)\}\Bigr|^-_{\p B_j}\right|\leq C|\tau_1-\tau_2|\frac{1}{\sqrt\ep}\frac{1}{2\beta}O(1)+O(1).\eeq
Since $|t-1|\leq -\log t \mbox{ for all }0<t<1$,
$$\frac{\left|\tau_1-\tau_2\right|}{-\log\tau}
\leq\frac{\left|\tau_1-1\right|+\left|1-\tau_|\right|}{-\log \tau_1-\log\tau_2}
\leq 2.$$
Hence the normal derivative inside $B_1$ and $B_2$ is bounded. Similarly, we prove the tangential derivative is bounded.
\qed

Following the proof of Proposition \ref{cornor}, it can be easily shown the following.
\begin{prop}\label{cornor2}
Suppose $\tau_1,\tau_2>0$.
The normal derivative of $\Im\{q(\Bx;\beta,\tau_1,\tau_2)\}$ is remain bounded regardless of $\ep$ for $\Bx\in\RR^2\setminus\overline{B_1\cup B_2}$. For $\Bx\in B_1\cup B_2$, \begin{align*}
&\ds\pd{\Im\{q\}}{\nu}\bigg|_{\xi=c}=\frac{\cosh c+\cos\theta}{2r_*\sqrt\ep}\left(\tau_1+\tau_2+2\tau\right)\Re\left\{ P\bigr(e^{-|c|+i\theta};\beta\bigr)\right\}+O(1).
\end{align*}
The tangential derivative satisfies
\begin{align}&\pd{\Im\{q\}}{T}\bigg|_{\xi=c}= -\mbox{sgn}(c) \left(\tau_1+\tau_2+2\tau\right)\frac{\cosh c+\cos\theta}{2r_*\sqrt\ep}
\Re\left\{ P\bigr(e^{-\max(\xi_s,|c|)-i\theta)};\beta\bigr)\right\}+O(1),\quad\Bx\in\RR^2.\label{cortaneqn}
\end{align}
\end{prop}

\begin{cor}\label{asymp:1127} Suppose $k_1,k_2>1$ From Theorem \ref{main_thm_u_x}, Theorem \ref{main_thm_u_y}, Proposition \ref{cornor} and Proposition \ref{cornor2}, we have
$$\bigr|\nabla (u-H)\bigr|^+(\Bx)= |c_\mathbf{n}|\left(|\tau_1|+|\tau_2|+2\tau\right)\frac{\cosh\xi_j+\cos\theta}{2r_*\sqrt\ep} \left|\Re\{P\bigr(e^{-(\xi_j+i\theta)};\beta\bigr)\}\right|+O(1)$$
on $\p B_j$, $j=1,2$.
Especially at $\Bx_j$, which is the point on $\p B_j$ closest to the other disk, we have $\theta=0$ and
$$\bigr|\nabla (u-H)\bigr|^+(\Bx_j)=|c_\mathbf{n}|\left(|\tau_1|+|\tau_2|+2\tau\right)\frac{\cosh\xi_j+1}{2r_*\sqrt\ep} \left|\Re\{P\bigr(e^{-\xi_j};\beta\bigr)\}\right|+O(1).$$
The above asymptotic equations are still valid for $0<k_1,k_2<1$ if $\mathbf{n}$ is replaced by $\mathbf{t}$.
\end{cor}
\begin{remark}\label{remark1}
Applying Lemma \ref{at0}, it follows that there is a constant $C_1$ and $C_2$ such that
\begin{align}\label{accor1}\left|\nabla u\right|_+(\Bx_j)&\geq \frac{C_1r_*(\nabla H(\Bp)\cdot\mathbf{n})}{\sqrt{\ep}(\beta+1)},\quad j=1,2,\\\label{accor2}\left\|\nabla u\right\|_{L^\infty(\p B_j)}&\leq \frac{C_2r_*(\nabla H(\Bp)\cdot\mathbf{n})}{\sqrt{\ep}(\beta+1)},
\end{align}
where $\Bp$ is the middle point of the shortest line segment connecting $\p B_1$ and $\p B_2$.
Since
$\beta=\frac{r_*}{4\sqrt{\ep}}(1-\tau+O((1-\tau)^2)$,
$$\frac{r_*}{\sqrt\ep (\beta+1)}\approx\frac{4}{1-\tau+\frac{4}{r_*}\sqrt\ep}.$$
Recall that ${r_*}/\mbox{max}(r_1,r_2)\leq{1}/{r_*}\leq{r_*}/{\mbox{min}(r_1,r_2)}.$
Hence \eqnref{accor1} and \eqnref{accor2} are in accordance with the upper and lower bounds \eqnref{gblowup} and \eqnref{gblowup2}.
\end{remark}

\subsection{Extreme cases for $\beta$}
In this section we let us consider the case when $\beta$ is extremely small or large. Recall that $\beta$ in \eqnref{parameter_beta} can be an arbitrary positive number even when $\ln\tau$ is very small because the denominator is small parameter. For example,
$\beta =\frac{r_*}{4\sqrt{\ep}}(1-\tau+O((1-\tau)^2)\approx{\epsilon}^{1/4}$ if $k_j\approx\epsilon^{-3/4}$, and $\beta\approx\ep^{-1/4}$ if $k_j\approx\ep^{-1/4}$.

From the definition $P$ in \eqnref{PPhi}, it can be written as
$$P(z;\beta)=-\frac{1}{z+1}+\beta\Phi(-z,1,\beta).$$
Set $|z|<1$, namely, $z=-e^{-\xi+i\theta}$ with $\xi>0$. For $\beta\ll1$,
\begin{align}\label{betasmall}
\ds \Phi(z,1,\beta) &=\frac{1}{\beta}+\sum_{k=0}^{\infty} \frac{z^k}{k}-\beta\sum_{k=1}^\infty\frac{z^k}{k^2(1+\frac{\beta}{k})}=\frac{1}{\beta}-\log (1-z)+O(\beta)
\end{align}
and  $$\Re\{\Phi(z,1,\beta)\}=\frac{1}{\beta}-\ln\sqrt{2e^{-\xi}(\cosh \xi+\cos\theta )}+O(\beta).$$
If $\beta\gg1$, applying the integration by parts twice on \eqnref{Lerch_int}, we have
\beq\ds\label{betabig}
\Phi(z,1,\beta)
=\frac{1}{\beta}\frac{1}{1-z}+\frac{1}{\beta^2}\frac{-z}{(1-z)^2}+\frac{1}{\beta^3}\frac{r(z,\beta)}{(\cosh \xi+\cos\theta)^{\frac{3}{2}}},\eeq
with $|r(z,\beta)|\leq 32.$
It is worth to mention that there are more complete results on the asymptotic expansions of $\Phi(z,s,\beta)$ for large and small $\beta$ done by Ferreira and Lop\'{e}z \cite{FL}.

 Applying the asymptotic of the Lerch transcendent in \eqnref{betasmall} and \eqnref{betabig}, we have
\begin{align}\Re\{P(e^{-\xi_j-i\theta};\beta)\}&=\frac{1}{2}-\frac{\sinh \xi_j}{2(\cosh \xi_j+\cos\theta)}-\beta\ln|1+e^{-\xi_j-i\theta}|+O(\sqrt\ep)\nonumber\\\label{betasmallP}
&=\frac{1}{2}\Bigr(1-\beta\ln[2(\cosh \xi_j+\cos\theta)]\Bigr)+\frac{O(\sqrt\ep)}{\cosh\xi_j+\cos\theta},\quad \mbox{for }\beta=O(\epsilon^{1/4}),\end{align}
and
\begin{align*}
\Re\{P(e^{-\xi_j-i\theta};\beta)\}&=\frac{1}{\beta}\frac{e^{-\xi_j-i\theta}}{(e^{-\xi_j-i\theta}+1)^2}+\frac{O(\sqrt\ep)}{(\cosh\xi_j+\cos\theta)^{\frac{3}{2}}}\\
&=\frac{1}{\beta}\frac{1+\cosh\xi_j\cos\theta}{2(\cosh \xi_j+\cos\theta)^2}+\frac{O(\sqrt\ep)}{(\cosh\xi_j+\cos\theta)^{\frac{3}{2}}},\quad\mbox{for }\beta=O(\epsilon^{-1/4}).
\end{align*}

One can easily show that there is a constant $C$ independent of $\ep$ such that \beq\label{ubound:1126}|\nabla (u-H)(\Bx)|\leq C, \quad\mbox{for }\Bx\in\RR^2\setminus(B_1\cup B_2)\mbox{ satisfying }\{|\theta|>\pi-\sqrt\ep\}.\eeq
Note that $\{|\theta|>\pi-\sqrt\ep\}$ is a region whose distance from the touching point is bigger than a positive constant independent of $\ep$. Here $\sqrt\ep$ is not the optimal rate for the region where the gradient $\nabla(u-H)$ is bounded independently of $\ep$. For the case of $\beta=O(\ep^{1/4})$ we can prove the following lemma.
\begin{lemma}\label{lem:ep14}
For $\ep,k_1,k_2$ satisfying $\beta=O(\ep^{1/4})$, there is a constant $C$ independent of $C$ such that
$$|\nabla (u-H)(\Bx)|\leq C, \quad\mbox{for }\Bx\in\RR^2\setminus(B_1\cup B_2)\mbox{ satisfying }\{|\theta|>\pi-\ep^{1/4}\}.$$
\end{lemma}
\pf
If $|\theta|>\pi-\ep^{1/4}$, then $1+\cos\theta=O(\sqrt\ep)$. Hence
$({\cosh\xi_s+\cos\theta})/{\alpha}=O(1).$ From \eqnref{betasmallP}, we prove the lemma.
\qed
\subsection{Non-uniform Convergence of $u_k$ to $u_\infty$}\label{sub:nonuniform}

For the perfectly conducting case, $k_1=k_2=\infty$, the conductivity problem becomes
\beq\label{u_inf:eqn}
\quad \left\{
\begin{array}{ll}
\ds\Delta u=0 \quad& \mbox{in } \mathbb{R}^2 \setminus \overline{B_1 \cup B_2}, \\
\ds u= \mbox{constant}\quad& \mbox{on }\p B_j, j=1,2,\\
\nm
\ds \int_{\p B_j} \pd{u}{\nu^{(j)}} ~ds=0,\ \quad& j=1,2,\\
\ds u(\Bx)-H(\Bx)=O(|\Bx|^{-1}) \quad&\mbox{as } |\Bx| \to \infty,
\end{array}
\right.
\end{equation}
where $\nu^{(j)}$ is outward normal to $\p B_j$. For this case, we can represent $q(\beta;0,1,1)$ in terms of linear combination of $\xi$ and $\theta$. Using \eqnref{eqnc1c2} and \eqnref{assump}, we have
\begin{align*}
 q(\Bx;0,1,1)
&=-\log\left(1+e^{-2\xi_1-(\xi+i\theta)}\right)+\log\left( 1+e^{-2\xi_2+(\xi+i\theta)}\right) \\
&=-\log\left(1+e^{-2\xi_1} \frac{\alpha-z}{\alpha+z}\right)+\log\left( 1+e^{-2\xi_2}\frac{\alpha+z}{\alpha-z}\right).
\end{align*}
Hence
\begin{align}
q(\Bx;0,1,1)&=-\log\left( \frac{(1-e^{-2\xi_1})(z+\alpha \coth{\xi_1})}{\alpha+z} \right)+\log\left( \frac{(1-e^{-2\xi_2})(z+\alpha \coth{\xi_2})}{\alpha-z} \right)\nonumber\\
&=\log\left(\frac{\alpha+z}{\alpha-z}\right)+\log\left(\frac{z+\tilde{\Bc}_2}{z-\tilde{\Bc}_1}\right)+
\log\frac{1-e^{2\xi_2}}{1-e^{-2\xi_1}}\nonumber\\
&=\xi+i\theta + b ,\label{q011}
\end{align}
where $(\Re\{\tilde{\Bc}_j\},\Im\{\tilde{\Bc}_j\})=\Bc_j$, $j=1.2$, and
$\nabla b$ is uniformly bounded independently of $\epsilon$.

Let us denote $u_k$ the solution to \eqnref{cond_eqn} with $k_1=k_2=k$ and $u_\infty$ to \eqnref{u_inf:eqn}. In Appendix of \cite{BLY}, it was shown the weak $H^1$-convergence of $u_k$ to $u_\infty$ for fixed $\ep$ when the background domain is bounded and inclusions are of strictly convex shape. For circular inclusions we have the follows.
\begin{lemma}\label{W1inf}
Let $H(\Bx)=x$ and $u_\infty$ be the solution to \eqnref{u_inf:eqn}. Then, for fixed $\epsilon>0$, we have
$$u_k\rightarrow u_\infty\quad\mbox{in }W^{1,\infty}(\RR^2),\quad \mbox{as }k\rightarrow \infty.$$
\end{lemma}
\pf
For $H(\Bx)=x$, as in the same way in the proof of Lemma \ref{lem:bipolar}, one can see that $u_\infty=x+\Re\{U'\}$ where $U'$ is defined as the following:
\begin{equation} \notag
 \ds U'(x,y)=C'+\begin{cases}
 \ds \sum_{n=1}^{\infty} \left(A'_n e^{n (\xi+i\theta)} + B'_n e^{-n( \xi+i\theta)}\right) ,\quad &x\in\mathbb{R}^2 \setminus (B_1 \cup B_2)\\
 \ds \sum_{n=1}^\infty \left(A'_n e^{n(\xi+i\theta)}+B'_ne^{n(2\xi_1+\xi-i\theta)}\right),& x\in B_1\\
 \ds  \sum_{n=1}^\infty\left(A'_ne^{n(2\xi_2-\xi+i\theta)}+B'_n e^{-n( \xi+i\theta)}\right),& x\in B_2\\
 \end{cases}
\end{equation}
where $C'=-\sum_{n=1}^{\infty} (A_n' + B_n' )\cos n\pi$,
\begin{align*}
\ds A'_n&=\frac{2\alpha(-1)^n}{ e^{2n(\xi_1+\xi_2)}-1}\left(-e^{2n\xi_1}-1\right),\mbox{ and }
\ds B'_n=\frac{2\alpha(-1)^n}{ e^{2n(\xi_1+\xi_2)}-1}\left(1+e^{2n\xi_2}\right).
\end{align*}

By the maximum principle, without loss of generality, it is enough to consider $\Bx \in \p B_1 \cup \p B_2$. Moreover, we may consider only $\Bx \in\p B_1$. To estimate  $\|u_k-u_\infty\|_{L^\infty(\p B_1)}$ for $k$ goes to infinity, we need to do the following quantities:
$$
\sum_{n=1}^{\infty} (A_n-A'_n)e^{-n \xi_1} \cos n\theta, \quad
\sum_{n=1}^{\infty} (B_n-B'_n)e^{n \xi_1} \cos n\theta, \quad \mbox{and}\quad
(C-C').
$$
We compute
\begin{align}
|A_n-A'_n|&=
\left| \frac{\tau_1^{-1}{e^{2n\xi_1}}+1}{\tau^{-1} e^{2n(\xi_1+\xi_2)}-1}-\frac{e^{2n\xi_1}+1}{e^{2n(\xi_1+\xi_2)}-1}\right|\nonumber\\
&\leq M \left( |\tau_1^{-1}-\tau^{-1}|+|1-\tau_1^{-1}|+|1-\tau^{-1}|\right)\frac{e^{2n(2\xi_1+\xi_2)}}{(e^{2n(\xi_1+\xi_2)}-1)^2}\nonumber\\
&\leq M\frac{1}{k}\frac{e^{2n(2\xi_1+\xi_2)}}{(e^{2n(\xi_1+\xi_2)}-1)^2}. \label{An_An}
\end{align}
Therefore we have
$
|\sum_{n=1}^{\infty} (A_n-A'_n)e^{-n \xi_1} \cos n\theta| \leq {M}/{k}.
$
Similarly, we can prove the convergence for $\sum_{n=1}^{\infty} (B_n-B'_n)e^{n \xi_1} \cos n\theta$ and $(C-C')$.

Let us now consider $\|\nabla(u-u_\infty)\|_{L^\infty(\p B_1)}$. For $\Bx\in \p B_1$, we have
\begin{align*}
|\nabla(u_k-u_\infty)(\Bx)| &\leq C   \sum_{n=1}^{\infty}n\left(|A_n-A'_n|e^{-n\xi_1} + |B_n-B'_n|e^{n\xi_1} \right)\leq \frac{M}{k}.
\end{align*}
This completes the proof.
\qed

\begin{lemma}
Suppose $1/k=O(\epsilon^{\frac{3}{4}})$ and $1/k\gg \ep$. There is a positive constant $C$ independent of $\ep$ such that
$$\|\nabla (u_k-u_\infty)(\Bx_j)\|\geq \frac{C}{k\ep}, \ j=1,2,$$
where  $H(\Bx)=x$ and $\Bx_j$ is the point on $\p B_j$ closest to the other disk.
\end{lemma}

\pf
From the definition, $\beta=O(\ep^{1/4})$. Applying Lemma \ref{W1inf}, \eqnref{betasmallP} and \eqnref{charbup}, 
\beq\label{inftycompare}
\frac{\p}{\p \nu}(u_k- u_\infty)\Bigr|^+_{\p B_j} = c_\mathbf{n}\frac{\cosh \xi_j+\cos\theta}{r_*\sqrt\ep}\beta\ln[2(\cosh \xi_j+\cos\theta)]+O(1).\eeq
At $\Bx_j$, where $\theta=0$, $\frac{\p}{\p \nu}(u_k- u_\infty)$ is order of magnitude $1/(k\ep)$.
\qed

The above lemma says that the convergence in Lemma \ref{W1inf} is not uniform in terms of $\ep$. There is a practical implication of it in computing the electric field. The perfectly conducting boundary condition gives a good approximation if $\ep$ is bigger than $1/k$, where $|\nabla u_k|$ in the proximity of objects is as big as $\ep^{-1/2}$ and $|\nabla (u_k-u_\infty)|$ is bounded.
However, as $\ep\ll 1/k$, $|\nabla (u_k-u_\infty)|$ becomes as big as $1/(k\ep)$ while $|\nabla u_k|$ is still of magnitude $\ep^{-1/2}$. Hence the $L^\infty$-error in the computation of electric field can be arbitrary large.

\begin{figure}[h!]
\begin{center}
\epsfig{figure=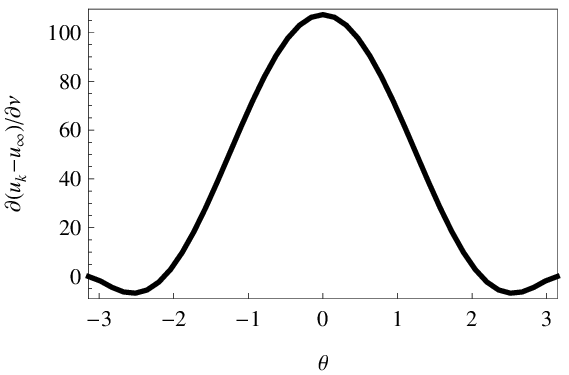,height=3.7cm}\hskip .2cm
\epsfig{figure=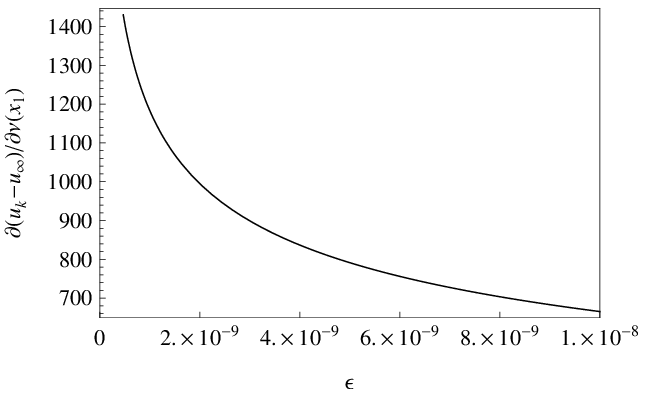, height=3.7cm}
\end{center}
\vskip -.5cm\caption{The left figure is the graph of $\p(u_k-u_\infty)/\p\nu\bigr|^+_{\p B_1}$ when $k_1=k_2=\epsilon^{-3/4}$ and $\epsilon=10^{-5}$. The normal derivative of $u_\infty$ is approximately 800. The right is the graph of $\p(u_k-u_\infty)/\p\nu\bigr|^+_{\p B_1}$ at $\Bx_1$ in terms of $\ep$.}
\label{fig_nonuniform}
\end{figure}

Let us visualize the non-uniform convergence with an example.
We set $H(\Bx)=x$, $r_1=2, r_2=3$ and $k_1=k_2=\epsilon^{-3/4}$. The centers of inclusions are located such that \eqnref{assump} is satisfied. In the first of Figure \ref{fig_nonuniform}, we plot the singular term of $\p(u_k-u_\infty)/\p\nu\bigr|^+_{\p B_1}$  in \eqnref{inftycompare} for $\epsilon=10^{-5}$ and $k_1=k_2=\epsilon^{-3/4} \approx 5623$. Figure \ref{fig_nonuniform} clearly shows that the difference diverges as $\epsilon$ decreases.

Since the harmonic conjugate of the solution $u$ of \eqnref{cond_eqn} is the solution to \eqnref{cond_eqn} with $k_j$ replaced by $1/k_j$, $j=1,2$,
we have the similar non-uniform convergence of the gradient of $u$ for the almost insulating case.
\section{Boundary Value Problem}\label{bvp}
Let $\Omega$ be a bounded domain in $\RR^2$ with $\mathcal{C}^2$-boundary containing two circular conducting inclusions $B_1$ and $B_2$. We assume that the inclusions are located away from $\p \Om$ and the distance between them is $\epsilon$. We consider the following boundary value problem:
\beq\label{cond_eqn_bounded}
\quad \left\{
\begin{array}{ll}
\ds\nabla\cdot\sigma\nabla u=0\quad&\mbox{in }\Om, \\
\ds \pd{u}{\nu}\Bigr|_{\p\Om}=g&(g\in L^2_0(\Om)),
\end{array}
\right.
\end{equation}
where $\nu$ is the outward unit normal to $\p \Om$.

For a bounded domain $B$ with $\mathcal{C}^2$ boundary, define the single- and double layer potentials as
\begin{align*}\Scal_B[\varphi](\Bx)&=\frac{1}{2\pi}\int_{\p B}\ln|\Bx-\By|\varphi(\By)d\sigma(\Bx),\quad \Bx\in \RR^2,\\
\Dcal_B[\varphi](\Bx) &= -\frac{1}{2\pi}\int_{\p B}\frac{\la\Bx-\By,\nu (\By)}{|\Bx-\By|^2}\varphi(\By)d\sigma(\By),\quad\Bx\in\RR^2\setminus\p B.\end{align*}
Following the same procedure as in the section 3 of \cite{KLY} to approximate $u$ by series of solutions to the free space solution, we get the following theorem.

\begin{theorem}
Suppose $k_1,k_2> 1$. The solution $u$ to \eqnref{cond_eqn_bounded} satisfies
\beq\nonumber
 u(\Bx)= c_\mathbf{n}\Re\{q(\Bx;\beta,\tau_1,\tau_2)\} + u_r(\Bx),
\eeq
where $c_\mathbf{n}$ is defined with 
\beq\label{def:H1012}H(\Bx)=-\Scal_\Om [g](\Bx)+\Dcal_\Om[u|_{\p\Om}](\Bx),\quad\Bx\in\Om\eeq and $|\nabla u_r(\Bx)|$ is uniformly bounded for $\ep$ .
\end{theorem}

\begin{theorem}
Suppose $0<k_1,k_2< 1$. The solution $u$ to \eqnref{cond_eqn_bounded} satisfies
\beq\nonumber
 u(\Bx)= c_\mathbf{t}\Im\{q(\Bx;\beta,-\tau_1,-\tau_2)\}+u_r(\Bx),
\eeq
where $c_\mathbf{t}$ is defined with $H$ in \eqnref{def:H1012} and $|\nabla u_r(\Bx)|$ is uniformly bounded for $\ep$ .
\end{theorem}

The characterization of the singular term of $u$ finds a very good application in the computation of electric fields. Computation of the gradient $\nabla u$ (or $u$) in the presence of adjacent inclusions with large or small conductivity value is a challenging problem due to the blow-up feature of the electric field.
Modifying the algorithm in \cite{KLY}, where the electric field in the presence of inclusions with extreme conductivities (perfectly conducting or insulating) was computed using $h$, to use $q$ in the place of $h$, one can accurately compute $u$ for the inclusions of arbitrary constant conductivity.

\section{Numerical Illustration}
In this section we demonstrate the main results for some examples. We compare the gradient of the solution $u$ to \eqnref{cond_eqn} with that of the singular term which is derived in Theorem \ref{main_thm_u_x} and Theorem \ref{main_thm_u_y}.
For notational convenience, let us denote the singular terms in Proposition \ref{cornoreqn} and Proposition \ref{cortaneqn} as follows:
\begin{align*}
Q_{\p B_1}(\theta)
&= r_*\left(|\tau_1|+|\tau_2|+2\tau\right)\frac{\cosh\xi_1+\cos\theta}{2\sqrt\ep}\Re\left\{ P\bigr(e^{-(\xi_1+i\theta)};\beta\bigr)\right\}.
\end{align*} Outside inclusions, only the normal derivative of $u$ blows-up in case of highly conducting inclusions while the tangential derivative does in case of almost insulating case.

For all examples, $r_1=3$, $r_2=2$, and the centers of them are located satisfying \eqnref{assump}. Since the graph shows the similar behavior either on $\p B_1$ or on $\p B_2$, we plot graphs only on $\p B_1$.\\

\noindent\textbf{Data Acquisition}
 To compute $u$, we use the series expansion in Lemma \ref{lem:bipolar}. Using \eqnref{nor_bipolar} and \eqnref{tan_bipolar}, both the normal and the tangential derivative of $u$ can be represented as series as well as $u$. It is worth to remark the difficulty in these computation for small $\ep$.
Note that $e^{-2n(\xi_1+\xi_2)}$ term is involved in Lemma \ref{lem:bipolar} and, hence, the cost in numerical computation becomes very high for small $\epsilon>0$. For instance, in Example 1, we evaluate the summation for $n \leq 5*10^3$ to compute within a tolerance $10^{-8}$. If we change $\epsilon$ to $10^{-6}$, then $10$-times more summation is required for the same tolerance.

The singular function $Q_{\p B_1}$ can be represented in terms of Lerch transcendent function $\Phi$. Using its analytic properties (such as integral representation), very efficient numerical algorithms are already implemented in many commercial softwares. In this paper, it is used 'LerchPhi' built-in function in Mathematica. Moreover, for the case of extremely small and large $\beta$, the Lerch transcendent $\Phi$ can be approximated in terms of elementary functions. Therefore the computational cost of computing the singular terms in $\nabla u$ (or in $u$) becomes much smaller than the one based on its series representation.
\smallskip

\begin{figure}[h!]
\begin{center}
\epsfig{figure=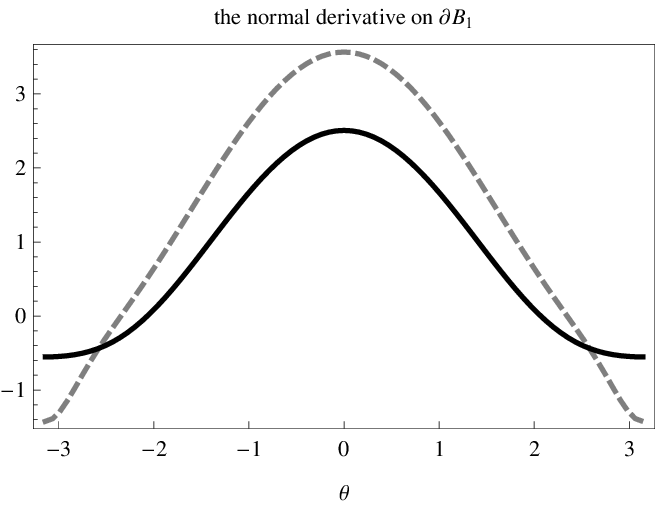,width=4cm}\hskip .2cm
\epsfig{figure=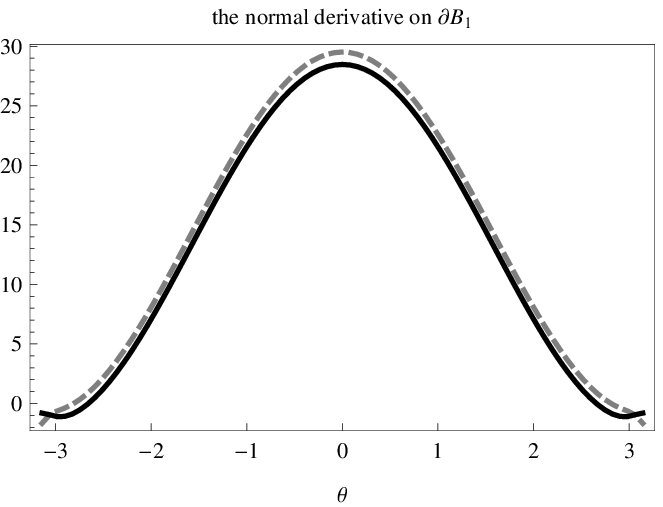,width=4cm} \hskip .2cm
\epsfig{figure=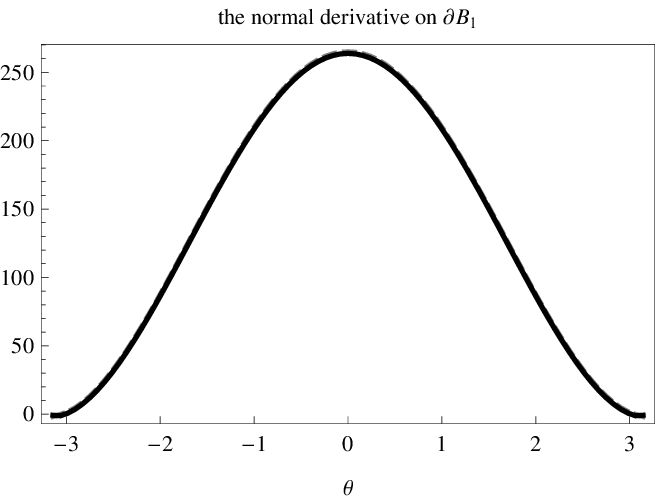,width=4cm}
\end{center}
\vskip -.5cm\caption{Highly conducting case. $k_1=1500,k_2=1200$, and $\epsilon$ is $0.5,0.01,0.0001$ from left to right.
The dashed gray line represents $\p (u-H)/\p \nu\bigr|^+_{\p B_1}$ while the black line does $Q_{\p B_1}$.}
\label{fig_ep_decr}
\end{figure}

\noindent{\bf Example 1}.
In Figure \ref{fig_ep_decr}, we show the numerical illustration for normal derivative for highly conducting case ($\nabla u$ of almost insulating case with $H(\Bx)=y$ haves the same feature). Fixing the conductivities of inclusions ($k_1=1500$ and $k_2=1200$), we change  $\ep$ ($\epsilon=0.5,0.01,0.0001$). The background potential is $H(\Bx)=x$. We plot the graph of the normal derivative $\p (u-H)/\p \nu\bigr|^+_{\p B_1}$ and $Q_{\p B_1}$ in gray and black, respectively. It shows that the difference between two terms are remained almost unchanged while the magnitude of them increases as $\ep$ decreases.
\smallskip

\begin{figure}[h!]
\begin{center}
\epsfig{figure=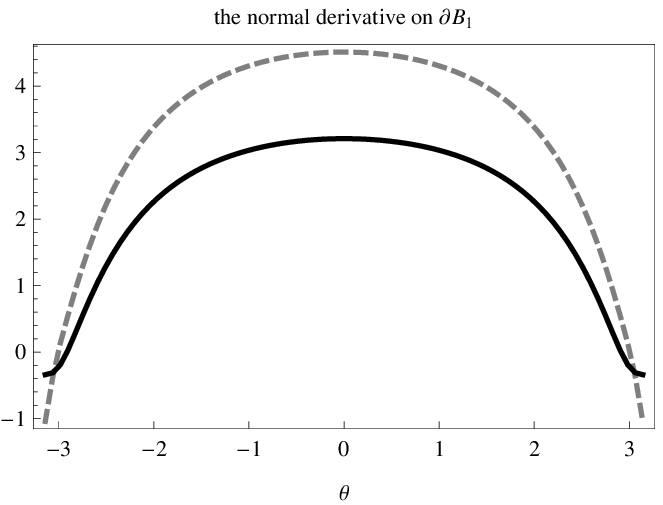,width=4cm}\hskip .2cm
\epsfig{figure=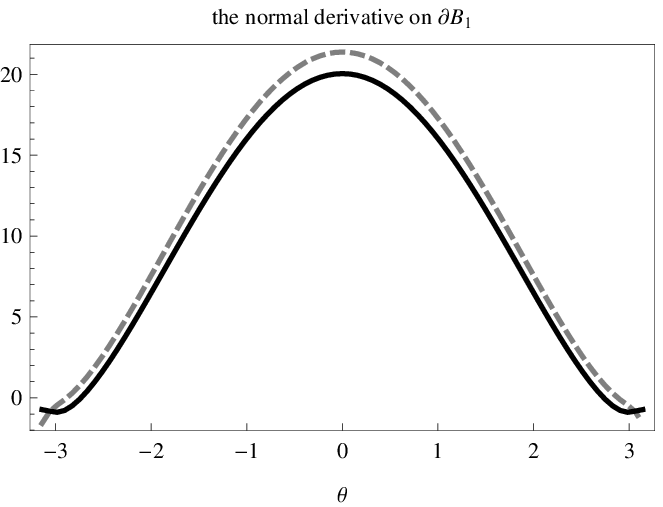,width=4cm} \hskip .2cm
\epsfig{figure=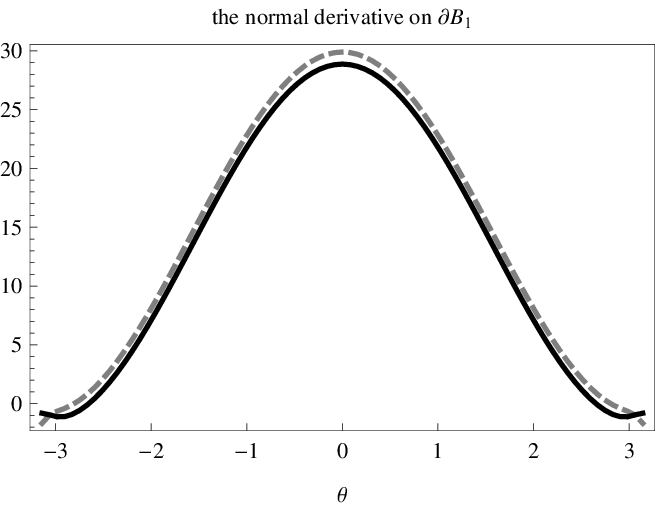,width=4cm}
\end{center}
\vskip -.5cm\caption{The distance $\epsilon$ is fixed to be 0.01, and conductivities are $(k_1,k_2)$ is $(7,5),(70,50),(7000,5000)$ from left to right.}
\label{fig_cond_incr}
\end{figure}

\noindent{\bf Example 2}. In this example, we set $\epsilon=0.01$ and $H(\Bx)=x$. The conductivities are $(k_1,k_2)=(7,5),(70,50),(7000,5000)$.
In Figure \ref{fig_cond_incr}, we plot $\p (u-H)/\p \nu\bigr|^+_{\p B_1}$ and $Q_{\p B_1}$ in gray and black, respectively. The difference between two terms are remained almost unchanged.

\smallskip

\noindent{\bf Example 3}. In this example, we draw the equipotential lines for the singular term in  $u$ and $|\nabla u|$ outside inclusions when the circular inclusions are highly conducting or almost insulating. More precisely, we set $(k_1, k_2)=(100,200),(0.03,0.02)$ and fix $\epsilon=0.1$. The background potential $H(\Bx)$ is  $x$ for highly conducting inclusions and $y$ for almost insulating ones.

\begin{figure}[h!]
\begin{center}
\epsfig{figure=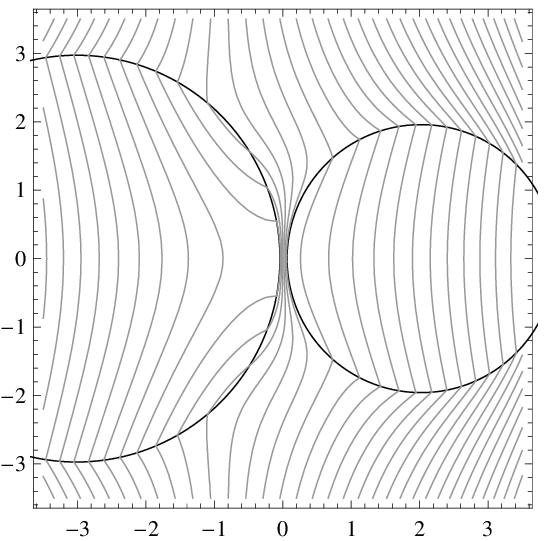,width=3.7cm} \hskip .4cm
\epsfig{figure=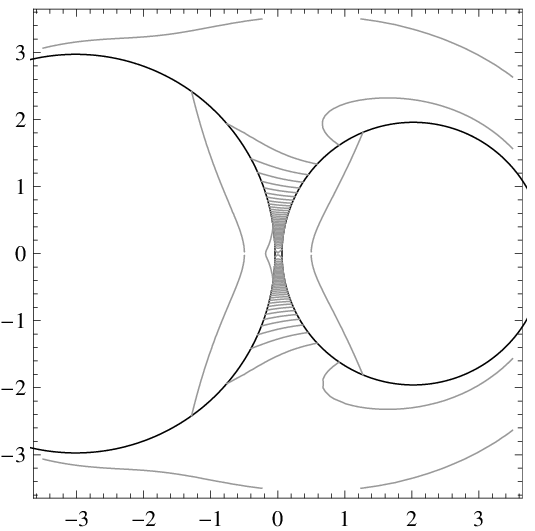,width=3.7cm}\\
\epsfig{figure=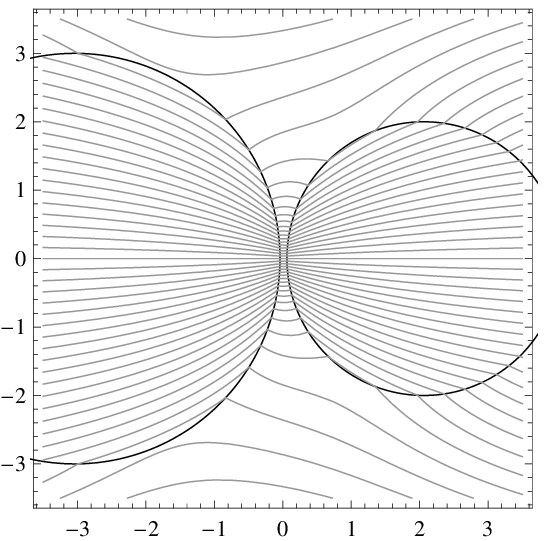,width=3.7cm} \hskip .4cm
\epsfig{figure=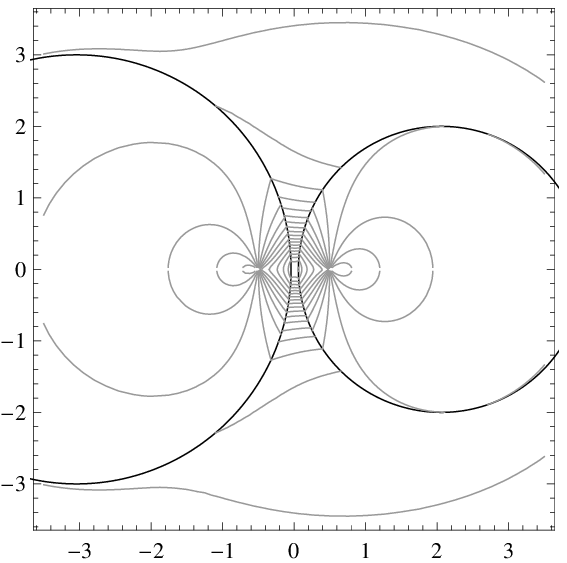,width=3.7cm}
\end{center}
\vskip -.5cm\caption{Contour plot of the singular term of $u$, the first column, and of $|\nabla u|$, the second column. Inclusions are either highly conducting, the first row, or almost insulating, the second row.}
\label{contour_singulaighk}
\end{figure}

In Figure \ref{contour_singulaighk}, we show contour plot of $\Re\{q\}$ and $\bigr|\Re\{\nabla q\}\bigr|$ which are singular terms of $u$ and $|\nabla u|$, respectively. For both cases, in the narrow region between the inclusions, $|\nabla u|$ changes fast in $y$-direction but slowly in $x$-direction which is in accordance with Proposition \ref{cornor}. Away from the touching region, $|\nabla u|$ changes slowly, see Lemma \ref{lem:ep14}. The contour of $|\nabla u|$ is continuous across the boundary of inclusions for almost insulating case while it is discontinuous for highly conducting, and $|\nabla u|$ blows-up inside inclusions for almost insulating case while it is bounded for highly conducting case as $\ep$ goes to zero.
%

\end{document}